%% file: main.tex
\documentclass[nonacm]{acmart}
\input{preamble}

\setcopyright{acmcopyright}
\acmYear{2023}
\copyrightyear{2023}
\title{Assessing the Ability of ChatGPT to Screen Articles for Systematic Reviews}

\author{Eugene Syriani}
\orcid{0000-0001-6527-1651}
\affiliation{
  \institution{DIRO, Université de Montréal}
  \city{Montréal}
  \country{Canada}
}
\email{syriani@iro.umontreal.ca}

\author{Istvan David}
\orcid{0000-0002-4870-8433}
\affiliation{
  \institution{DIRO, Université de Montréal}
  \city{Montréal}
  \country{Canada}
}
\email{istvan.david@umontreal.ca}

\author{Gauransh Kumar}
\affiliation{
  \institution{DIRO, Université de Montréal}
  \city{Montréal}
  \country{Canada}
}
\email{gauranshk21@gmail.com}

\begin{document}

\darkmodecontrol{}

\input{sections/abstract}
\maketitle

\input{sections/introduction}
\input{sections/relatedwork}
\input{sections/method}

\input{sections/threats}
\input{sections/results}
\input{sections/discussion}
\input{sections/conclusion}

\input{sections/acknowledgement}

\bibliographystyle{ACM-Reference-Format}
\bibliography{bib/bibliography}

\clearpage

\input{sections/appendix}

\end{document}

%% file: preamble.tex
\input{preamble/packages}
\input{preamble/commands}
\input{preamble/settings}
\input{preamble/listings}

%% file: preamble/packages.tex
\usepackage[utf8]{inputenc}
\usepackage{ifthen}
\usepackage{hyperref}

\usepackage{soul}
\usepackage[normalem]{ulem}
\usepackage{graphicx}

\usepackage{epsf,picinpar}
\usepackage{varioref}
\usepackage{fdsymbol}

\usepackage{enumerate}
\usepackage[shortlabels]{enumitem}
\usepackage{geometry}

\usepackage{pifont}
\usepackage{fancyhdr}
\usepackage[pscoord]{eso-pic}

\usepackage{pdfcomment}
\usepackage[framemethod=TikZ]{mdframed}
\usepackage{listings}
\usepackage{courier}
\usepackage{multirow}

\usepackage{tikz}
\usetikzlibrary{positioning,decorations.pathreplacing,fit}
\usepackage{subcaption}
\usepackage{tabularray}

%% file: preamble/commands.tex
\newcommand{\secref}[1]{Sec.~\ref{#1}}

\newcommand{\figref}[1]{Figure~\ref{#1}}
\newcommand{\tabref}[1]{Table~\ref{#1}}

\newcommand{\lstref}[1]{Listing~\ref{#1}}
\newcommand{\appref}[1]{Appendix~\ref{#1}}
\newcommand{\eg}{e.g.,~}
\newcommand{\ie}{i.e.,~}

\newcommand{\stdess}[1]{%
    \ifthenelse{\boolean{showannotations}}%
    {\textcolor{magenta}{\textbf{STANDARD-ESSENTIAL}:~{#1}}}%
    {}%
}

\newcommand{\stddes}[1]{%
    \ifthenelse{\boolean{showannotations}}%
    {\textcolor{green}{\textbf{STANDARD-DESIRABLE}:~{#1}}}%
    {}%
}

\newcommand{\stdex}[1]{%
    \ifthenelse{\boolean{showannotations}}%
    {\textcolor{green}{\textbf{STANDARD-EXTRAORDINARY}:~{#1}}}%
    {}%
}

\newcommand{\todo}[1]{%
    \ifthenelse{\boolean{showannotations}}%
    {\ifthenelse{\equal{#1}{}}{\textcolor{red}{TODO}}{\textcolor{red}{TODO:~{#1}}}}%
    {}%
}

\newcommand{\assignedto}[1]{%
    \ifthenelse{\boolean{showannotations}}%
    {\textbf{\noindent\ding{46}\textcolor{white}{~\colorbox{\assignementcolor}{Assigned to:}}~\textcolor{\assignementcolor}{#1}\\}%
    }
    {}
}

\newcommand{\rem}[1]{%
    \ifthenelse{\boolean{showannotations}}%
    {\textcolor{\oldtextcolor}{\st{#1}}}%
    {}%
}

\newcommand\add[1]{%
    \ifthenelse{\boolean{showannotations}}%
    {\textcolor{\newtextcolor}{{#1}}}%
    {#1}%
}

\newcommand\rep[2]{%
    \ifthenelse{\boolean{showannotations}}%
    {\rem{#1}~\add{#2}}%
    {#2}%
}

\newcommand{\acmchecklist}[3]{%
    \ifthenelse{\boolean{showacmchecklist}}%
    {\pdfmarkupcomment[color=green]{#1}{#2: #3}}%
    {#1}%
}

\newcommand{\darkmodecontrol}{%
    \ifthenelse{\boolean{darkmode}}%
    {\pagecolor{black!20}
    \color[rgb]{0,0,0} 
    }%
    {}%
}

\newcommand{\parameter}[1]{{\small{\texttt{#1}}}}

\newcommand\encircle[1]{%
  \tikz[baseline=-3px]
    \node (X) [draw=black, fill=circleyellow, shape=circle, inner sep=0,scale=0.75,text=black] {\strut #1};}

\newcommand{\tikzmark}[2][]{%
  \tikz[remember picture,overlay,baseline=-.5ex] \node[#1] (#2) {};%
}

\newcommand*{\drawBrace}[4][0pt]{%
    \node[draw=none, fit={(#2) (#3)}, inner sep=0pt] (rectg) {};%
    \draw [decoration={brace,amplitude=0.3em},decorate,very thick,orange]%
      ([xshift=#1]rectg.north east) --%
      coordinate[right=0.9em, midway] (#2#3)
      ([xshift=#1]rectg.south east);%
    \node[right=-0.8em of #2#3] (#2#3-comment) {\color{orange} \textbf{#4}};
}%

\newenvironment{conclusionframe}[1]
  {\mdfsetup{
    frametitle={\colorbox{white}{\space#1\space}},
    innertopmargin=2pt,
    frametitleaboveskip=-\ht\strutbox,
    frametitlealignment=\center
    }
  \begin{mdframed}%
  }
  {\end{mdframed}}

%% file: preamble/settings.tex
\newboolean{showannotations}
\setboolean{showannotations}{true} 

\newboolean{shownames}
\setboolean{shownames}{true}

\newboolean{showacmchecklist}
\setboolean{showacmchecklist}{false} 

\newboolean{darkmode}
\setboolean{darkmode}{false} 

\newcommand{\newtextcolor}{blue}
\newcommand{\oldtextcolor}{red}
\newcommand{\assignementcolor}{orange}
\definecolor{highlightcolor}{rgb}{.99, 1, .0}
\sethlcolor{highlightcolor}
\definecolor{circleyellow}{HTML}{ffff99}

\newcommand{\framecolor}{gray!5}
\mdfsetup{%
   middlelinecolor=black,
   middlelinewidth=0.75pt,
   backgroundcolor=\framecolor,
   roundcorner=2pt,
   skipabove=0.5\baselineskip,
   frametitlealignment=\center
   }

\makeatletter
\newcommand\notsotiny{\@setfontsize\notsotiny{6.75}{7.3}}
\makeatother

%% file: preamble/listings.tex
\definecolor{darkgreen}{rgb}{0,0.6,0}
\definecolor{gray}{rgb}{0.5,0.5,0.5}
\definecolor{mauve}{rgb}{0.58,0,0.82}
\definecolor{gray}{rgb}{0.4,0.4,0.4}
\definecolor{darkblue}{rgb}{0.0,0.0,0.6}
\definecolor{lightblue}{rgb}{0.0,0.0,0.9}
\definecolor{cyan}{rgb}{0.0,0.6,0.6}
\definecolor{darkred}{rgb}{0.6,0.0,0.0}
\definecolor{armygreen}{rgb}{0.29, 0.33, 0.13}
\definecolor{antiquefuchsia}{rgb}{0.57, 0.36, 0.51}
\definecolor{coolblack}{rgb}{0.0, 0.18, 0.39}

\lstdefinelanguage{prompt}
{
    comment=[l]{--},
    morekeywords={READ},
    keywordstyle=\color{white},
    morestring=[s][\color{darkred}\bfseries]{\{}{\}},
    morestring=[s][\color{darkred}\bfseries]{[}{]},
    morestring=[s][\color{mauve}\bfseries]{(}{.)},
    extendedchars=true
}

%% file: sections/abstract.tex
\begin{abstract}
  By organizing knowledge within a research field, Systematic Reviews (SR) provide valuable leads to steer research. Evidence suggests that SRs have become first-class artifacts in software engineering. However, the tedious manual effort associated with the screening phase of SRs renders these studies a costly and error-prone endeavor. While screening has traditionally been considered not amenable to automation, the advent of generative AI-driven chatbots, backed with large language models is set to disrupt the field. In this report, we propose an approach to leverage these novel technological developments for automating the screening of SRs. We assess the consistency, classification performance, and generalizability of ChatGPT in screening articles for SRs and compare these figures with those of traditional classifiers used in SR automation. Our results indicate that ChatGPT is a viable option to automate the SR processes, but requires careful considerations from developers when integrating ChatGPT into their SR tools.
\end{abstract}

\keywords{
generative AI,
GPT,
empirical research,
large language model,
literature review,
LLM,
mapping study,
review,
screening,
survey
}

%% file: sections/introduction.tex
\section{Introduction}\label{sec:introduction}

Systematic Reviews (SRs) are a scholarly method for synthesizing and organizing knowledge from primary studies within a specific research field. As a secondary study, an SR aims to ``identify, analyze, and interpret all available evidence related to a specific research question''~\cite{kitchenham2007guidelines}. These reviews document the state-of-the-art and provide a foundation for academic scholars to guide their research toward impactful directions.

In the field of software engineering, the number of published SRs has been steadily increasing, with 1$\,$723 recorded on the DBLP computer science bibliography website.\footnote{
  \url{https://dblp.uni-trier.de/search/publ?q=systematic\%20(review\%7Cmapping)} -- Accessed on 2023-05-25.
}
Despite their importance, conducting an SR can be challenging and labor-intensive. Among the various phases of an SR, the screening process, which involves selecting relevant scientific articles for inclusion, has been reported as the most time-consuming~\cite{borah2017analysis}. It is also a primary source of errors in building the article corpus due to its manual nature, introducing threats to internal validity such as fatigue, attrition, and researcher biases~\cite{mallett2012benefits}.

To address these challenges, researchers commonly employ strategies such as assigning multiple reviewers for each article, restricting screening to titles and abstracts, and introducing a validation step by a senior reviewer~\cite{kitchenham2007guidelines}. While these practices help address some of the challenges associated with manual screening, they do not scale well with large article corpora, ultimately making human performance a bottleneck in the SR process.
Given that working with corpora of thousands of articles is not uncommon, screening remains a critical problem in SRs. Experts have identified article screening as one of the most significant barriers~\cite{alzubidy2017vision}, leading to increased costs associated with conducting SRs.

The need for computer-aided automation or reduction of manual screening tasks holds significant value in SRs. Several studies~\cite{felizardo2020automating,tsafnat2014systematic,jonnalagadda2015automating} have acknowledged this need, leading to the development of various software tools that assist reviewers throughout the review process. However, most of these tools do not automate the screening phase, as it has traditionally been considered challenging to automate~\cite{felizardo2020automating}. State-of-the-art screening automation tools typically rely on ranking articles based on the likelihood of inclusion~\cite{martinez2008facilitating}. Nevertheless, determining the optimal stopping point for screening articles remains unclear, as empirical studies have shown significant variations across different SRs~\cite{marshall2019systematic}.

With the emergence of large language models (LLMs), such as GPT~\cite{floridi2020gpt}, the automation of screening activities has become feasible. LLMs are AI models that have been pre-trained on vast amounts of textual data, enabling them to capture extensive knowledge that can be utilized, among other things, for classifying articles within a corpus.

This study aims to assess \textbf{whether ChatGPT can be used to assist in screening articles in an SR}.
At the time of writing, OpenAI's GPT family of models constitutes the largest LLM, which has been widely utilized in various domains beyond software engineering, including public health care~\cite{biswas2023role}, climate research~\cite{biswas2023potential}, and creative writing~\cite{mcgee2023what}.
To achieve our objective, we conducted an exploratory study using ChatGPT\footnote{
  \url{https://openai.com/chatgpt/}
} as the LLM service in April--June 2023, and addressed the following research questions:

\begin{enumerate}[\bfseries{RQ}1.]
   
    \item \textit{How \textbf{consistent} are the decisions made by ChatGPT?}
    We investigate the consistency of ChatGPT's decisions regarding specific articles within an SR.
    Consistency is an important quality to mitigate threats to construct validity when relying on ChatGPT.

    \item \emph{How does the \textbf{classification performance} of ChatGPT compare to traditional classifiers used in SR?}
    Classification performance encompasses performance metrics typically used to evaluate classifiers and metrics specific to the screening problem.
    Assessing classification performance is crucial for understanding the potential of ChatGPT in supporting SRs.

    \item \textit{How \textbf{generalizable} are the decisions made by ChatGPT?}
    The development of automation tools for SR is justified if the solution can generalize across a representative range of problems.
    We investigate whether the classification performance of ChatGPT is similar in different SRs conducted on different topics.

    Developing automation for SR tools is justified only if the developed solution generalizes over a representative class of problems. Here, we investigate if the classification performance of ChatGPT is similar in multiple SRs conducted on different topics.

\end{enumerate}

\paragraph{Results}
Our results show that an LLM can perform as well as machine learning methods traditionally used for automating SR activities. However, ChatGPT achieves this without additional training. Our results have important implications on the automation of SRs as they show that LLMs exhibit superior performance over state-of-the-art machine learning techniques in SRs and have a realistic chance to revolutionize SR automation.

\paragraph{Structure}
The rest of this report is structured as follows.
In \secref{sec:relatedwork}, we review the related work.
In \secref{sec:method}, we discuss the methods used in our study.
In \secref{sec:threats}, we discuss the threats to validity.
In \secref{sec:results}, we present the results and address the research questions.
In \secref{sec:discussion}, we discuss the results.
Finally, in \secref{sec:conclusion}, we conclude our report.

%% file: sections/relatedwork.tex
\section{Related work}\label{sec:relatedwork}

In the past two decades, researchers from various domains have explored different approaches to reduce the screening effort in SRs. Some of these efforts involve improving well-established protocols.
For instance, \citet{kosar2018systematic} propose a variation of the protocol for SRs in software engineering introduced initially by \citet{kitchenham2007guidelines}. Other efforts focus on automation through machine learning techniques. \citet{marshall2019systematic}, for example, propose active learning methods for ranking articles to be screened.

In a recent systematic literature review on automation in SRs, \citet{Dinter2021} identified 41 relevant studies, primarily in the fields of medicine and software engineering. Their findings reveal that the most commonly used machine learning models for automation are Support Vector Machines (SVM) \cite{joachims1998text} and Bayesian Networks, while Bag of Words and Term Frequency-Inverse Document Frequency (TF-IDF) are the most popular natural language processing representation techniques. Their review highlights that no study has yet investigated the use of deep neural network models specifically for the screening phase of an SR. One of the main challenges van Dinter et al. identified is the issue of imbalanced datasets, with many excluded articles dominating the distribution. This skewed distribution often leads classifiers to maximize the overall accuracy. Additionally, the dependence on current automation techniques based on machine learning approaches requires manual crafting and fine-tuning of features for each particular domain or dataset.

Several tools focus on automating the screening phase using machine learning to classify or rank articles. With the recent popularity of LLMs, only a few have explored their potential to aid researchers in conducting SRs. However, none of them have specifically investigated the use of LLMs in the screening phase of an SR. In the following, we delve into these existing works, providing insights into their applications, most of which have been conducted on medical datasets.

\subsection{A posteriori reduction of screening work}\label{sec:screening-support}

Most screening automation techniques require a labeled dataset to be trained on
Therefore, the reduced work effort can only be known after the screening has been already conducted manually, defeating the very purpose of automation.

\citet{martinez2008facilitating} propose a technique to prioritize the corpus of articles by ranking them from most to least likely to be included.
It uses a generic text retrieval search engine and a classifier that re-ranks the retrieved articles.
To evaluate their approach, they use the Work Saved over Sampling (WSS) metric \cite{cohen2006reducing} and show that, on some datasets, the technique would have saved over 50\% of the effort.

\citet{cohen2006reducing} train a boosted perceptron-based classifier to predict when new articles should be added to SRs on drug class efficacy for the treatment of disease. Their approach could theoretically reduce the number of abstracts that need to be screened by up to 68\% while maintaining a recall of 95\% to the eligible citations.
\citet{matwin2010new} employ a factorized version of the complement Naive Bayes classifier to maximize recall when it was too low with their original algorithm.
\citet{ji2017using} utilize an information retrieval technique that establishes relationships and ontology-based semantics of the articles.
\citet{amarjeet2018fp} explore the Fuzzy-Pareto dominance-driven artificial bee colony algorithm for multi-objective software module clustering, which can be used in articles clustering too.

\subsection{Ranking articles by active learning}\label{sec:screening-ai}

Active learning is a machine learning technique in which the learning agent is allowed to choose the training data from which it learns and is allowed to query an oracle to label previously unlabeled instances~\cite{settles2009active}.
Prioritizing articles that are more likely to be included helps the informed human to proceed at a higher pace when making decisions about inclusions. Conversely, prioritizing articles the algorithm is less sure about allows it to learn faster and rank the remaining articles with higher confidence.
Despite this apt idea, such screening techniques are still not widely used due to their limited generalizability~\cite{Khabsa2016} and limited efficiency~\cite{Miwa2014}.

\citet{wallace2010semi} implement active learning based article screening by Support Vector Machines (SVM) using the SVMLIB library.\footnote{
  \url{https://www.csie.ntu.edu.tw/~cjlin/libsvm/}
}
Abstrackr is an online tool notably relying on this technique~\cite{wallace2012deploying}.
Unfortunately, our experiments with Abstrackr on a corpus from software engineering were not convincing. 
A possible explanation we found is that Abstrackr is tuned for articles in medicine, where abstracts are much better structured than those in software engineering. Typically, abstracts of medical articles are succinct excerpts of the full paper, disclosing experimental details, results, and conclusions. The relatively simplistic abstracts of software engineering articles might simply lack essential information for Abstrackr to work effectively.

\citet{marshall2019systematic} follow an active learning variant based on certainty.
The classifier is continuously trained on manually screened articles.
It then predicts the probability of relevance for all unseen articles and reorders them by presenting to the reviewer those most relevant first. The cycle continues as the reviewer screens papers and the model re-ranks the remaining ones. When uncertainty sampling is used, papers predicted with the least certainty are presented first to improve the models' accuracy more efficiently.
The trick is to determine how many positive examples will suffice to achieve good predictive performance. A conservative heuristic is about half of the dataset, but this should be determined empirically.
They list the following tools that support this process: Abstrackr, Colandr, EPPI reviewer, SWIFT-Review, and RobotAnalyst. The latter two also group articles by similar topics.

Van de Schoot et al.~\cite{vandeschoot2021open} present ASReview an open-source machine-learning-aided pipeline with active learning for SRs.
It allows the user to choose from multiple machine learning models: Naive Bayes, SVM, deep neural network, logistic regression, LSTM-base, LSTM-pool, and Random Forest.
They offer various feature extraction models: embedding with Inverse Document Frequency (IDF) or TF-IDF, Sentence BERT~\cite{reimers2019sentence}, Doc2Vec~\cite{Le2014}, and long short-term memory networks (LSTM).
To evaluate the performance, they use WSS and the number of relevant references found after having screened the first 10\% of the records.

\citet{ferdinands2020active} show that a Naive Bayes classifier with TF-IDF performs better than SVM for their four datasets. However, they notice that dataset characteristics significantly affect the performance of the classifier.

\citet{rozanc2021chapter} propose automating the screening task for systematic mapping studies. They have tool support that needs to heavily be configured iteratively by the human for each SR.
They employ a text statistic analysis technique to count the occurrence of important words, iteratively defined decision rules, and a screening pilot. However, their approach is applicable only if the full text of the articles is taken into account in the screening, which is not the common way \cite{kitchenham2007guidelines}.

\subsection{LLMs and their application in SRs}\label{sec:llm4sr}

The recent advances in LLMs---especially popularized with the infatuated AI-driven chatbot ChatGPT---have instigated a revolutionary paradigm shift in software engineering and many disciplines \cite{zhou2023comprehensive}.
ChatGPT is a Generative Pre-trained Transformer that employs an auto-regressive language model trained on large datasets with billions of tokens from CommonCrawl, Wikipedia, and other publicly available text sources.
It relies on a deep neural network with a transformer architecture to estimate the conditional probability of a sequence of tokens given a context. Through reinforcement learning from human feedback, ChatGPT is continuously improving its performance.

Pre-training LLMs of natural language can be achieved by using an unlabeled textual corpus. The main limitation of the techniques discussed in \secref{sec:screening-ai} is that they require to be trained on a large set of labeled articles either beforehand or during the screening process. Thus, using ChatGPT to reduce the reviewer's workload in screening articles without training it specifically on the corpus of the SR seems like a promising solution.

Fine-tuning LLM, such as ChatGPT can be achieved in two ways: hyperparameter tuning and prompt engineering.
Its main tuneable hyperparameter is the {\small\texttt{temperature}}, a value between 0 and 1 to control the diversity of its response to a prompt.
The prompt, i.e., the context and instruction the user provides to the bot is paramount to be well-designed.
There are many prompt engineering approaches \cite{Liu2023}, like zero-shot, N-shots, or chain-of-thought (CoT).
Zero-shot prompt means that the user explains the task without providing any labeled example.
N-shots mean that at least $N$ labeled example solutions of the task are provided with the prompt.
CoT means that the prompt includes reasoning steps along with the instructions.

LLMs have experienced a steep adoption curve since early 2023, and have seen applications beyond text generation.
We found only two articles focusing on using ChatGPT to support SRs. However, neither of them has addressed the task of screening articles.
\citet{wang2023can} have studied the use of ChatGPT to automatically formulate search queries to retrieve articles.
They experimented on a dataset from PubMed for a medical SR with 70 titles and abstracts from a standard test benchmark \cite{alharbi2018retrieving}.
They obtain high precision but low recall on this corpus. One problem they faced is that ChatGPT generates different queries even if the same prompt is used, which impacts its effectiveness and reproducibility. They conclude that it is not clear that ChatGPT can effectively be used to generate SR search queries.
\citet{waseem2023conducting} propose a method for Human-Bot collaboration while conducting SRs.
However, screening is manual in their approach and ChatGPT can be used as a decision support system to assist reviewers, but not to replace them.

%% file: sections/method.tex
\section{Methodology}\label{sec:method}

We followed the Data Science method~\cite{ralph2021empirical}, relying on a data-centric analysis method in our study.
This empirical method is the most appropriate for our purpose given the large quantitative dataset, we use to answer the research questions and the data-intensive analysis of the quality metrics identified in the research questions.

\begin{figure*}[h!]
    \centering
    \includegraphics[trim={0.75cm 3.5cm 0.25cm 2.25cm},clip,width=0.9\linewidth]{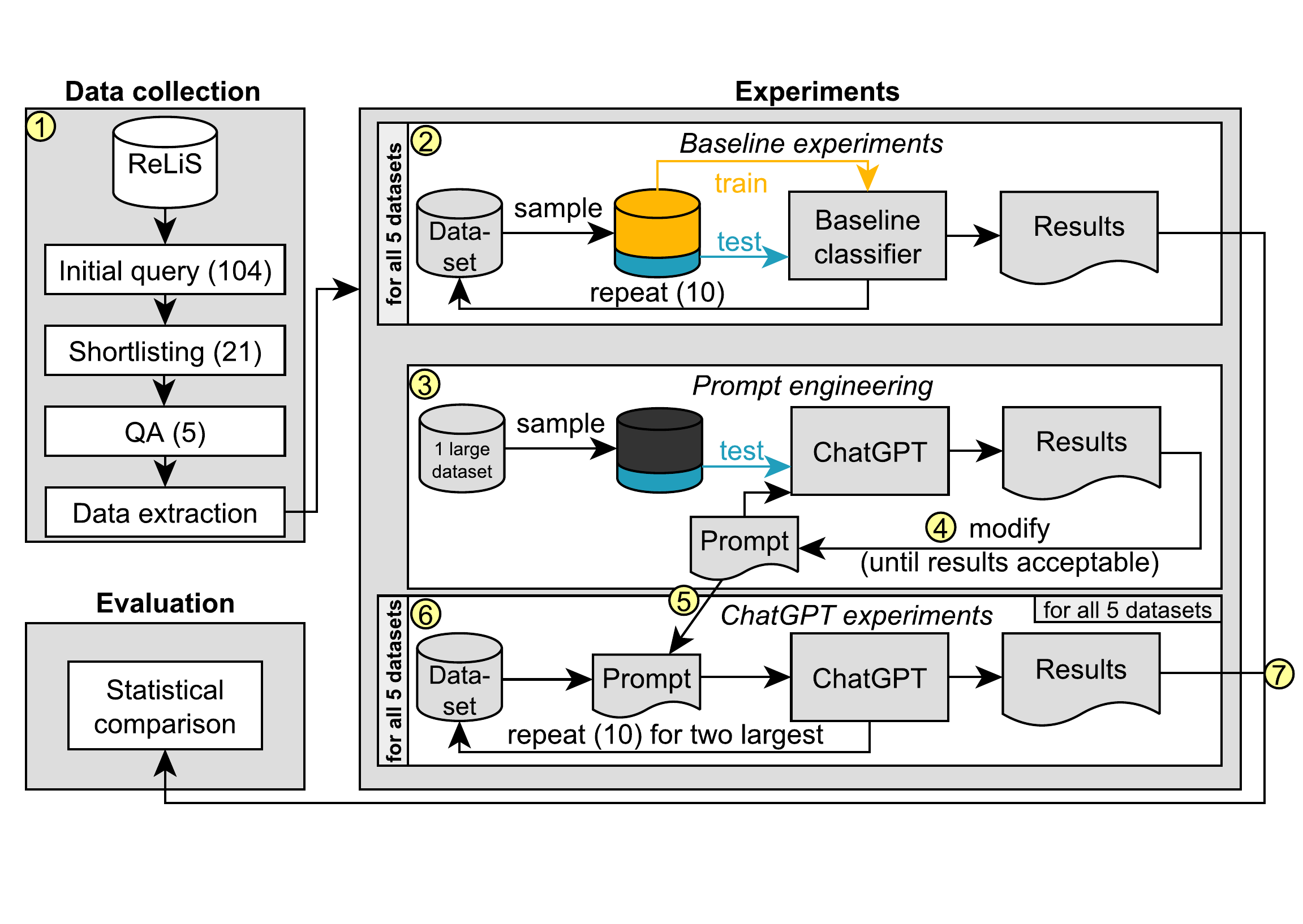}
    \caption{Overall study design}
    \label{fig:workflow}
\end{figure*}

\figref{fig:workflow} shows the overall design of the study.
After collecting the data \encircle{1}, we conduct a three-phase experiment to answer the research questions.
First, we set up a baseline by conducting experiments with machine learning models used in the literature to classify articles in SRs \encircle{2}.
Then, we conduct experiments with ChatGPT by first engineering the prompt \encircle{3}--\encircle{4}, and using the final prompt \encircle{5} to evaluate articles via ChatGPT \encircle{6}.
Finally, we evaluate the performance of the baselines and ChatGPT by comparing their statistical properties \encircle{7}.

\input{sections/method/data}
\input{sections/method/problem}
\input{sections/method/experiments}

%% file: sections/method/data.tex
\subsection{Data collection}\label{sec:datacollection}
 We now elaborate on the details of the data collection strategy.

\subsubsection{Data source}

ReLiS~\cite{bigendako2018modeling} is a cloud-based tool for planning, conducting, and reporting SRs\footnote{\url{https://relis.iro.umontreal.ca/}}.
Although most SRs publish their data in a replication package or appendix, replication packages contain only the final corpus of included articles. ReLiS stores the whole history of SR projects, including information about articles excluded during the screening phases, the exclusion criteria that were applied, and whether the decision was unanimous or required the resolution of a disagreement among the reviewers.
In essence, ReLiS projects provide corpora that have been carefully labeled by highly qualified experts and, therefore, can be considered as the ground truth to evaluate the decisions an AI would make about the inclusion or exclusion of articles in the corpora.

We extract the datasets used in our experiments through careful pre-processing and filtering steps explained below.

\subsubsection{Collecting datasets}\label{sec:datasets}
As of April 2023, ReLiS contained a total of 104 SR projects.

We shortlist ReLiS projects that define a screening phase. These are the projects that are likely real SR projects (i.e., not test, abandoned, or empty projects).
We queried the SQL database behind ReLiS to fully automate the retrieval of the shortlist, which resulted in 21 projects.

We then manually inspected the shortlisted projects to select those containing real and meaningful data of proper quality. Our selection criteria are that (1) the project is either concluded and led to a scholarly publication or (2) the project is still in progress and we can verify that it is conducted in a systematic manner. We consider that criterion (1) is a good indication that the scientific community found the work sound, and by extension, we can assume that the associated ReLiS project contains a corpus that has been labeled correctly.

We manually searched digital libraries (Google Scholar, Scopus, and DBLP) for potential published articles corresponding to the identified ReLiS projects based on the project users, contact information, topic, and date. Some of the publications we found explicitly mentioned using ReLiS, such as \citet{barisic2022multi}, which increased our confidence. We also confirmed the correspondence between the publication and ReLiS project by contacting the authors of the publications. Eventually, we are able to identify five relevant ReLiS projects: three concluded projects with an associated publication and two ongoing projects with multiple rounds of screening and a substantial number of articles. Furthermore, we investigated the included articles in each project to ensure the correctness of the decisions. We found one project where the articles included did not match the topic of the review. After confirming with the authors, we discarded this project.
\tabref{tab:datasets} lists the five ReLiS projects we finally selected to form the datasets of our experiments\footnote{Refer to the Ethics statement at the end of this paper.}.

\subsubsection{Data extraction}\label{sec:extraction}

To use the selected projects in the experiments, we extract screening information from them. The dataset of a particular project contains all screening phases: all screened articles, including snowballing phases. A data record in each dataset represents a single screened article with the following data:
\begin{description}
  \item[Project:] Identifying the dataset this article comes from. Note that screening an article in one project may have included it and excluded it in another project for a different topic.
  \item[Key:] Unique identifier of the article within a project.
  \item[Title:] Short, textual ASCII string.
  \item[Abstract:] Full abstract as an ASCII string.
  \item[DOI:] Digital object identifier of the article.
  \item[Decision:] Binary value recording if the article was included or excluded by the reviewers. This is the ground truth.
  \item[Exclusion criteria:] Upon exclusion, we record its reason. In case of multiple reviewers, we list all exclusion criteria.
  \item[Reviewers:] Number of ReLiS users who reviewed this article.
  \item[Conflict:] Binary value recording if the decision was a result of a conflict eventually resolved among the reviewers.
\end{description}

If an article has been screened by multiple reviewers (which is typically the case in SRs), the same article will appear in the data set multiple times. From this, we can also reconstruct whether the article has been included/excluded unanimously or whether a conflict had to be resolved among reviewers. We record this information for later analysis purposes.

We filter data records to retain only those that have all the above data. For example, we discard screened articles without an abstract recorded in ReLiS or articles that are still pending reviewer decisions. Furthermore, we exclude duplicate entries within projects. Duplicate detection and removal, although an important step in SRs, is only semi-automated, and duplicate entries might still exist in the corpora.

\subsubsection{Characteristics of the datasets}\label{sec:charcteristics}

\begin{figure}[t]
    \centering
    \includegraphics[width=0.75\linewidth]{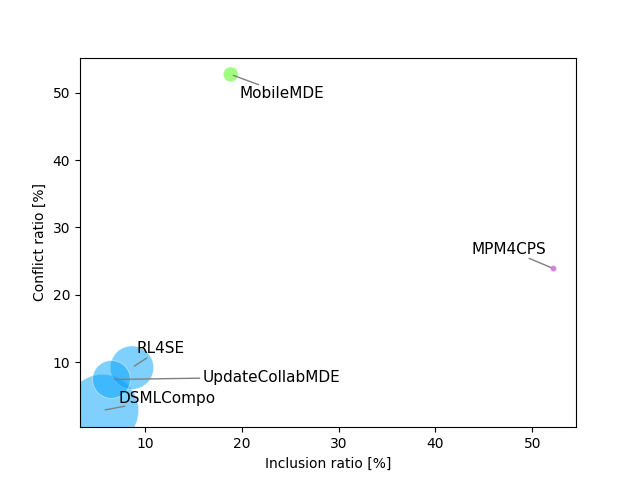}
    \caption{Inclusion ratio and conflict ratio of the dataset. Size proportional to corpus size.}
    \label{fig:bubble}
\end{figure}

\tabref{tab:datasets} and \figref{fig:bubble} present the characteristics of the five datasets after pre-processing.

\input{tables/datasets}

The datasets vary in size from 205 to $2\,683$ total records. The ratio of included articles ranges from 7\% to 52\%, with a median of 9\% and an average of 18\%. This is typical for SRs that are imbalanced with a predominance of excluded articles.
The three largest datasets have similar inclusion ratios. Note that UpdateCollabMDE is a systematic update study, meaning that all articles are collected through forward snowballing, not using a carefully designed query like the other SRs.
The inclusion ratio of the two smaller datasets varies significantly.
MPM4CPS includes more than half of the articles, whereas MobileMDE includes nearly 19\% of the articles, more than twice the ratio of the largest datasets.

The number of articles that had conflicting decisions between reviewers has a similar distribution, with an average of 19\%.
The datasets also differ from each other with the conflict ratios.
DSMLCompo has very few conflicts reported (3\%).
RL4SE and UpdateCollabMDE have similar conflict ratios around 8\%.
MPM4CPS and MobileMDE reported conflicts for a quarter and half of the articles respectively.

As evidenced by \figref{fig:bubble}, we work with three larger datasets (DSMLCompo, RL4SE, UpdateCollabMDE -- \textcolor{blue}{blue} cluster), each positioned in the 0--10\% range of inclusion ratio and conflict ratio, which are usual numbers for SRs. We also have two datasets with special profiles: MobileMDE (\textcolor{green}{green} cluster of one) has an atypical high conflict ratio and MPM4CPS (\textcolor{purple}{purple} cluster of one) has an atypical high inclusion ratio. The former profile is caused by repeated diverging decisions among the reviewers which may indicate an ambiguous scope or exclusion criteria of the SR. In the latter profile, we are faced with a balanced dataset of included and excluded articles. This may occur when a corpus is prefiltered.

Although ReLiS hosts SR projects with topics in different disciplines, our final dataset only contains SRs related to software engineering. Therefore, the chosen datasets vary in size, balance between inclusions and exclusions, conflicting decisions, and topics in software engineering.

%% file: tables/datasets.tex
\begin{table*}[t]
  \caption{The datasets used in our experiments. (Topic descriptions avaialable in \appref{app:datasets}.)}
  \label{tab:datasets}
  {\footnotesize
  \begin{tabular}{llrrrrrp{0.25\linewidth}}
    \toprule
    \textbf{Project} & \textbf{Publication} & \textbf{Size} & \textbf{Included} & \textbf{Excluded} & \textbf{Conflicts} & \textbf{Reviewers} & \textbf{Project title} \\
    \toprule
    DSMLCompo       & \textit{In progress} & $2\,683$ & $150~(5.6\%)$ & $2\,533$ & $76~(2.8\%)$ & $4$ & Domain-specific modeling language composition   \\
    MobileMDE       & \citet{brunschwig2022modelling} & $292$ & $55~(18.8\%)$ & $237$ & $154~(52.7\%)$ & $3$ & Modeling on mobile devices  \\
    MPM4CPS         & \citet{barisic2022multi} & $205$ & $107~(52.2\%)$ & $98$ & $49~(23.9\%)$ & $2$ & Multi-paradigm modeling of cyber-physical systems  \\
    RL4SE           & \textit{In progress} & $1\,089$ & $94~(8.6\%)$ & $995$ & $100~(9.2\%)$ & $6$ & Reinforcement learning for software engineering  \\
    UpdateCollabMDE & \citet{david2021collaborative} & $875$ & $57~(6.5\%)$ & $818$ & $65~(7.4\%)$ & $3$ & Collaborative modeling \\ \midrule
    \textbf{Total} & & $\mathbf{5\,222}$  & $\mathbf{467}$ & $\mathbf{4\,755}$ & $\mathbf{473}$ & &   \\
    \bottomrule
  \end{tabular}
  }
\end{table*}

%% file: sections/method/problem.tex
\subsection{Problem formulation}\label{sec:problem}

Given an article $a=\left\{\mathit{features}\right\}$ with a set of features and a ground truth decision $d \in D$ to include or exclude the article from an SR, the problem of screening the article is to define a classifier $c$ with $\hat{d}=c(a) \in D$, where $\hat{d}$ is the decision output by the classifier.
In this work, the features we consider are the title and the abstract of the article, as well as the topic of the SR.
This is typically the primary information available at the initial screening phase that is indexed by most, if not all, digital libraries.
The classifier is an AI model, such as an LLM (in our case: ChatGPT) or machine learning models fine-tuned for this task (\eg SVM).

Ideally, the classifier should take the same decision as the ground truth, \ie $\hat{d}=d$.
We consider the domain of decisions $D=\{0,1\}$ where $1$ means the decision is to include the study and $0$ to exclude it.
However, we should distinguish between the different situations when this is not the case.
Thus, we define an evaluator $E: D \cdot D \rightarrow \{TP, TN, FP, FN\}$ that populates the confusion matrix as follows:
\begin{equation}\label{eq:confusion}
  E(d,\hat{d}) =
  \begin{cases}\vspace{-.5em}
    \mathit{TP} & \mbox{if } d=1 \wedge \hat{d}=1  \\\vspace{-.5em}
    \mathit{TN} & \mbox{if } d=0 \wedge \hat{d}=0  \\\vspace{-.5em}
    \mathit{FP} & \mbox{if } d=0 \wedge \hat{d}=1  \\\vspace{-.5em}
    \mathit{FN} & \mbox{if } d=1 \wedge \hat{d}=0
  \end{cases}
\end{equation}
The values of $E$ represent the classifier decisions that correctly include (true positive -- \textit{TP}), correctly exclude (true negative -- \textit{TN}), incorrectly include (false positive -- \textit{FP}), and incorrectly exclude (false negative -- \textit{FN}) articles, respectively.

\subsection{Metrics}\label{sec:metrics}

The four values of $E$ are the primary metrics to assess the performance of the classifiers. However, given that the datasets vary in size and inclusion/exclusion ratio, we rely on metrics that are derived from them and serve as a basis of comparison on a $[0,1]$ ratio scale.

\subsubsection{Base metrics}
We include standard classifier metrics to ensure the classifier includes articles correctly:
\begin{description}
  \item[Precision] measures the ability to include only articles that should be included.
  \begin{equation}\label{eq:prec}
    \mathit{Prec}=\frac{TP}{TP+FP}
  \end{equation}
  \item[Recall] measures the ability to include all articles that should be included.
  \begin{equation}\label{eq:rec}
    \mathit{Rec}=\frac{TP}{TP+FN}
  \end{equation}
\end{description}

However, we are also interested in evaluating the decisions to exclude articles. The classifier should reduce the workload of reviewers by excluding studies that are trivial excludes, i.e., articles that are clearly outside the scope of the SR.
Thus, we include the equivalent metrics as above, tailored for exclusions:
\begin{description}
  \item[Negative predictive value (NPV)] measures the ability to exclude only articles that should be excluded. It is analogous to precision but for negative values.
  \begin{equation}\label{eq:npv}
    \mathit{NPV}=\frac{TN}{TN+FN}
  \end{equation}
  \item[Specificity] measures the ability to exclude all articles that should be excluded. It is analogous to recall but for negative values.
  \begin{equation}\label{eq:spec}
    \mathit{Spec}=\frac{TN}{TN+FP}
  \end{equation}
\end{description}

A successful classifier for screening should miss as few relevant articles as possible (maximize recall) and save time for the reviewers by removing as many irrelevant articles as possible (maximize NPV).\footnote{As explained in \secref{sec:relatedwork}, WSS is a frequently used non-standard metric to evaluate of automated screening tools that balance between high recall and sufficient NPV~\cite{kusa2023analysis}.
However, \citet{kusa2023analysis} have recently shown that, when WSS is normalized to a $[0,1]$ scale, $\mathit{WSS}=\mathit{Spec}$.
Therefore, we use specificity instead of WSS when reporting the results.}

\subsubsection{Metrics for imbalanced data}
The previous set of metrics considers the inclusion and exclusion decisions separately. For the screening task, it is important to consider both classes at the same time.
Moreover, the datasets to screen are usually imbalanced favoring the exclusion class: there are more articles to exclude than to include in SRs (see \tabref{tab:datasets}). Thus, there is a need for aggregated metrics able to deal with imbalanced datasets. We choose the three metrics below.
\begin{description}
  \item[Balanced accuracy] is used to capture the accuracy of deciding both inclusion and exclusion classes. It is better suited than the traditional accuracy metric for imbalanced classes. It corresponds to the area under the receiver operating characteristic curve (AUC) when only one run is available \cite{sokolova2006accuracy}.
  \begin{equation}\label{eq:bacc}
    bAcc = \frac{\mathit{Rec} + \mathit{Spec}}{2}
  \end{equation}
  \item[F2] Typically, we report the $F1$ score as a compact representation of precision and recall.
  However, it weights precision and recall equally.
  In our case, the classifier must strive to avoid excluding studies that should have been included. That is, recall must be as high as possible.
  Thus, following \citet{Chawla2008}, we consider the cost of getting false negatives twice as costly as getting false positives.
  Therefore, we find $F2$ to be a more suited F-score for our problem.
  \begin{equation}\label{eq:f2}
    F_2 = 5 \cdot \frac{\mathit{Prec} \cdot \mathit{Rec}}{4 \cdot \mathit{Prec} + \mathit{Rec}}
  \end{equation}
  \item[Matthews correlation coefficient (MCC)] balances the ability to classify all articles as included or excluded correctly. It is often used as the singular metric for imbalanced data \cite{Bekkar2013,Chicco2023} as it gives more realistic performance estimation of binary classifiers in such cases than, e.g., the commonly used AUC metric~\cite{lavazza2023reliability}.
  We use the normalized MCC measure to ensure values are in the $[0,1]$ scale. A value under $0.5$ indicates performance worse than random.
  \begin{equation}\label{eq:mcc}
    \mathit{MCC} = \frac{TP \cdot TN - FP \cdot FN}{2 \cdot \sqrt{(TP+FP)(TP+FN)(TN+FP)(TN+FN)}} + 0.5
  \end{equation}
\end{description}

\subsubsection{Metrics for consistency}
On top of the metrics above, RQ3 focuses on the robustness of the classifier's decisions: if it tends to always produce the same decision for the same features.
\begin{description}
  \item[Fleiss' Kappa] measures the inter-rater agreement. In our case, we consider each run of the classifier as an independent rater and assess the consistency of their decisions for each article. $p_e$ is the expected agreement across all articles of inclusion and exclusion decisions and $p_o$ is the observed agreement for each article. A value above 0.81 typically indicates an almost perfect agreement \cite{Fleiss2003}.
  \begin{equation}\label{eq:kappa}
    \kappa=\frac{p_o-p_e}{1-p_e}
  \end{equation}
\end{description}

\subsection{Classifiers for the baseline experiments}\label{sec:baselineexp}

In our experiments, we use \textbf{GPT version 3.5 Turbo} through the ChatGPT service as the representative state-of-the-art LLM.
To contextualize the results we obtain from the experiments with ChatGPT, we compare the results to representative baselines. Due to the relatively early stage of research on LLM, no baselines or benchmarks exists to uniformly evaluate LLM except for experimental solutions and works in progress, such as HELM~\cite{liang2022holistic}. To evaluate the results, we select four classifiers that are frequently encountered in similar problems, such as text classification and SR screening automation \cite{vandeschoot2021open}. We rely on the \textit{scikit-learn}~\cite{Pedregosa2011} Python library implementation of these classifiers.
\begin{description}[topsep=0pt,noitemsep,leftmargin=1em]
  \item[Logistic regression (LR)] A simple linear model that is also one of the most efficient ones when outcomes are separable by a linear plane~\cite{lavalley2008logistic}. This is exactly the case in our experiments due to the binary decision.
  \item[Random forest (RF)] is representative of an aggregating model. It consists of a set of decision trees trained on random subsets of features. It is particularly useful for classifying high dimensional noisy data, such as text~\cite{islam2019semantics}.
  \item[Complement Naive Bayes (CNB)] is frequently used baseline in text classification problems~\cite{xu2017bayesian}. This variant performs better on imbalanced data in which one class has substantially higher representation than the other~\cite{rennie2003tackling}, like in our case.
  \item[C-Support Vector Classification (SVC)] is a frequently used implementation of SVM in existing SR tools to rank the studies by the likelihood they should be included. Substantially reduces the need for labeled training instances~\cite{joachims1998text}.
\end{description}
We also implement a \textbf{random classifier (RAND)} that randomly assigns inclusion/exclusion decisions to papers. We use it to ensure that no classifier is performing worse than random. Otherwise, it cannot be used as-is, needs to be trained on a larger and more diversified dataset, or better tuned.

The four classifiers (LR, RF, CNB, and SVC) mentioned above need to be trained and tested using appropriately sampled data from a specific dataset. To ensure a robust evaluation, we followed the widely used 80:20 random split, where 80\% of the data was used for training and the remaining 20\% for testing. For each corpus in \tabref{tab:datasets}, we performed a randomized grid search to tune the hyperparameters of the classifiers~\cite{bergstra2012random}.
To train the classifiers, we performed a 5-fold repeated cross-validation on each dataset.
During cross-validation, we optimized the fitting process based on the F2 score, which allows us to strike a balance between minimizing false negatives and including the correct articles.
The selected hyperparameters for each dataset can be found in \appref{app:hyperparameters}.
In this process, each classifier is retrained specifically for each dataset to maximize its performance.

To represent the features of each article, we employed the Word2Vec algorithm, which utilizes a two-layer neural network to capture word associations from text~\cite{Mikolov2013}. Unlike TF-IDF, Word2Vec is a word embedding technique that effectively captures semantic meaning and word relationships. This aspect is particularly advantageous for our problem, as the textual features of each article are not extensive enough to rely solely on term frequency statistics. By leveraging Word2Vec, we can extract richer contextual information and enhance the representation of our article features.

Each experiment is conducted using appropriately randomized and sampled data sets to ensure the proper statistical power of the results.
As time performance is not relevant for our study, we conduct the experiments on \acmchecklist{regular office equipment}{essential}{discusses the hardware and software infrastructure used}. We developed a program in Python to orchestrate the overall experiment, including interfacing with ChatGPT through its API.

More detailed data is available in \appref{app:hyperparameters}.

%% file: sections/method/experiments.tex
\subsection{Prompt engineering for ChatGPT}\label{sec:expsetup}

The goal of this phase is to engineer a prompt that performs well and can be used in subsequent experiments with ChatGPT. We aim to engineer a prompt that can support screening in any SR irrespective of its scope or topic.
To identify the best prompt, we first experiment with the manual chat interface of ChatGPT and observe how modifications to the prompt meet our expectations or result in unexpected responses.
Once the approximate prompt is found, we automate the process. We change from manual experimentation through the GUI to automated queries to the API of ChatGPT. For this, we need to generate proper samples from the datasets and tune the hyperparameters of ChatGPT appropriately.

\paragraph{Sampling} To facilitate a rapid turnaround and keep our experiments with the prompt economically feasible, we sample smaller batches from the overall RL4SE data set. The samples have a size of 20--40 articles. To ensure statistical similarity between samples, we developed a script that randomly selects articles from the dataset given a specified ratio of inclusions and exclusions. The fixed ratio ensures statistical similarity and mitigates threats to internal validity, while random sampling improves statistical power.

\paragraph{ChatGPT hyperparameters} We set two important hyperparameters: \parameter{temperature} and \parameter{max\_tokens}. The former controls the randomness of the text generated by ChatGPT. Higher temperatures result in more variance in the generated text and perceived higher creativity. In our experiments, we require consistent responses and no creativity in the generated text. Therefore, we set \parameter{temperature} to $0$. The \parameter{max\_tokens} parameter controls the length of the generated text and forms the basis of incurred costs. We aim to keep the response short and standardized. Thus, we opt for one-word responses from ChatGPT: \parameter{Include} or \parameter{Exclude}. One token roughly equals 4 characters in English.\footnote{\href{https://help.openai.com/en/articles/4936856-what-are-tokens-and-how-to-count-them}{help.openai.com/en/articles/4936856-what-are-tokens-and-how-to-count-them}} To accommodate the two responses, both of length 7, we first set the \parameter{max\_tokens} to 2. However, we observed numerous cases when the response did not fit the token boundary. Thus, we increased \parameter{max\_tokens} to 3.

\paragraph{Final prompt template}
Eventually, we arrived at the template in \lstref{lst:prompt-template} as the best-performing prompt template that maximizes the F2 score.

\input{prompts/template}
The prompt template consists of the following parts.
\begin{description}[leftmargin=1em]
  \item[Context:] Describes the context of the query, i.e., conducting an SR (Line 1), informing ChatGPT about the topic (Line 2), and requesting a strong focus on the topic. This latter information is important in cases where adjacent topics might be undesirable, e.g., including articles that focus on software engineering for reinforcement learning rather than reinforcement learning for software engineering.
  The context has one parameter:
    \begin{itemize}[topsep=0pt,noitemsep,leftmargin=1em]
        \item \parameter{topic}: Brief description of the review. Required attribute. Exactly one topic has to be specified.
    \end{itemize}
  \item[Instructions:] Provides ChatGPT with instructions about the specific \textit{Task} at hand.
  Instructions have three parameters:
    \begin{itemize}[topsep=0pt, noitemsep,leftmargin=1em]
        \item \parameter{inputs} Name of the input fields (e.g., title, abstract, keywords). Required attribute. At least one input has to be specified.
        \item \parameter{include\_word}: The literal that is expected from ChatGPT upon suggesting to include an article.
        \item \parameter{exclude\_word}: The literal that is expected from ChatGPT upon suggesting to exclude an article.
    \end{itemize}
  \item[Task:] The specific article to decide the inclusion of. The article has two parameters, corresponding to the feature on which to produce a decision:
    \begin{itemize}[topsep=0pt, noitemsep,leftmargin=1em]
        \item \parameter{title}: Title of the article. Required attribute. Exactly one title has to be specified.
        \item \parameter{abstract}: Full verbatim abstract. Required attribute. Exactly one abstract has to be specified.
    \end{itemize}
\end{description}

We introduced a handful of manual optimization in the final prompts. First, to decrease the number of false negatives, we ask ChatGPT to \textit{be lenient} (Line 10 in \lstref{lst:prompt-template}). We observed that this instruction indeed decreased the number of false negatives. However, it came at the cost of an increased number of false positives. As a consequence, the recall increased but precision decreased. This is in line with the optimization priorities explained previously.
We also kept the number of tokens (the length of the prompts) minimal to improve cost efficiency.

Listing \ref{lst:prompt-rl4se-noex} shows an instance of the prompt template applied to the RL4SE dataset with the work of \citet{barriga2022parmorel} being asked to be screened by ChatGPT.
\input{prompts/full}

Not every SR project defines its topic and scope in one succinct sentence we could use for the \parameter{topic} parameter of the prompts. To conduct our experiments, we assign topic descriptions to projects that do not have one, following the four-step process below.
\begin{enumerate}[\bfseries1.,leftmargin=1.25em]
  \item \textbf{Determine the scope of SR} based on elements from the published paper or protocol of the SR, especially the goal, search strings, inclusion and exclusion criteria, and the overview of anticipated results.
  \item \textbf{Formulate scope} starting from the title of the paper or the ReLiS project and re-formulate it in a more precise sentence based on step 1.
  \item \textbf{Evaluate formulation} by asking ChatGPT whether it understands the scope, and verify the explanation it gives.
  \item \textbf{Refine formulation} by iterating (up to three times) over the topic formulation at step 3 until ChatGPT's explanation is satisfactory.
\end{enumerate}
The topic descriptions of the datasets are available in \appref{app:datasets}.

%% file: prompts/template.tex
\begin{figure*}[h!]
  \begin{lstlisting}[mathescape, escapechar=!, language={prompt}, caption = {Prompt template}, label = {lst:prompt-template}]
I am screening papers for a systematic literature review. !\tikzmark{startContext}!
The topic of the systematic review is {TOPIC}[1].
The study should focus exclusively on this topic. !\tikzmark{endContext}! 
    
Decide if the article should be included or excluded from the systematic review.!\tikzmark{startInstructions}!
I give the {INPUTS}[+] of the article as input.
Only answer {INCLUDE_WORD}[1] or {EXCLUDE_WORD}[1].
Be lenient. I prefer including papers by mistake rather than excluding them by mistake.!\tikzmark{endInstructions}!
    
Title: {TITLE}[1]!\tikzmark{startTask}!
Abstract: {ABSTRACT}[1]!\tikzmark{endTask}!
  \end{lstlisting}
  \begin{tikzpicture}[overlay, remember picture]
    \drawBrace{startContext}{endContext}{Context};
    \drawBrace{startInstructions}{endInstructions}{Instructions};
    \drawBrace{startTask}{endTask}{Task};
  \end{tikzpicture}
\end{figure*}

%% file: prompts/full.tex
\begin{figure*}[h!]
\begin{lstlisting}[language={prompt}, caption = {Prompt template applied to the RL4SE data set}, label = {lst:prompt-rl4se-noex}]
I am screening papers for a systematic literature review.
The topic of the systematic review is reinforcement learning for software engineering.
The study should focus exclusively on this topic.

Decide if the article should be included or excluded from the systematic review.
I give the title and abstract of the article as input.
Only answer Include or Exclude.
Be lenient. I prefer including papers by mistake rather than excluding them by mistake.

Title: PARMOREL: a framework for customizable model repair
Abstract: In model-driven software engineering, models are used in all phases of the development process. These models must hold a high quality since the implementation of the systems they represent relies on them. Several existing tools reduce the burden of manually dealing with issues that affect models' quality, such as syntax errors, model smells, and inadequate structures. However, these tools are often inflexible for customization and hard to extend. This paper presents a customizable and extensible model repair framework, PARMOREL, that enables users to deal with different issues in different types of models. The framework uses reinforcement learning to automatically find the best sequence of actions for repairing a broken model according to user preferences. As proof of concept, we repair syntactic errors in class diagrams taking into account a model distance metric and quality characteristics. In addition, we restore inter-model consistency between UML class and sequence diagrams while improving the coupling qualities of the sequence diagrams. Furthermore, we evaluate the approach on a large publicly available dataset and a set of real-world inspired models to show that PARMOREL can decide and pick the best solution to solve the issues present in the models to satisfy user preferences.

\end{lstlisting}
\end{figure*}

%% file: sections/threats.tex
\section{Threats to validity}\label{sec:threats}

Here, we review the main threats to the validity of the study and discuss how we mitigated them.

\subsection{Construct validity}
Choosing the measures of evaluation poses the most substantial threat to construct validity. In particular, the $F_\beta$ metric based on which we trained the baseline models and optimized our prompts is a result of arbitrarily choosing the $\beta$ value, i.e., the weight between recall and precision. To mitigate this threat, we followed community standards when choosing $\beta = 2$~\cite{brownlee2020tour}. We could have also used MCC or balanced accuracy to mitigate the imbalance between the inclusion and exclusion classes. Nevertheless, our results show that all three metrics have similar trends in our datasets.

The results of comparison with baselines might be artifacts of the training characteristics of classifiers we used, rather than meaningful observations about the superior performance of ChatGPT. To mitigate this threat, we used a grid search to tune the models as recommended by community standards~\cite{ralph2021empirical}.
Each model was retrained specifically for each dataset using cross-validation. Thus, the models are not meant to be used to screen any SR and are therefore biased towards each dataset. This is to say that we can consider the trained classical classifiers as ``good enough'' to establish a baseline when assessing the performance of ChatGPT.

We relied on plain text corpora that might contain editorial errors due to special characters and their encoding. For example, an en dash (``--'') might be encoded as ``\textbackslash endash'', and percentages might follow LaTeX conventions, e.g., ``25\textbackslash \%''. These errors might impact the performance of ChatGPT. We mitigated this threat by either removing problematic articles from the dataset or by applying meaningful clean-up transformations.

\subsection{Internal validity}
We used manually classified datasets in our experiments and specifically, to determine the performance metrics. Due to manual labor, these metrics are subject to threats to internal validity. To mitigate these issues, we selected datasets that are either associated with published peer-reviewed SR or are ongoing efforts in which the authors of the current paper are involved and can judge their quality.

\subsection{External validity}
Our study has sampled only SRs that are from the software engineering domain and have been conducted in the ReLiS SR tool. Furthermore, ChatGPT was the only LLM we evaluated. These choices pose threats to the external validity of the study, i.e., the generalizability of the results. On the one hand, our study was focusing only on the SE domain and generalizations to other domains require careful consideration of the target domain. On the other hand, we are reasonably confident that the main takeaways related to the accuracy of ChatGPT translate well to other LLMs of a similar kind.

%% file: sections/results.tex
\section{Results}\label{sec:results}

In this section, we present the results of the experiments directly addressing the research questions.

\input{sections/results/consistency}
\input{sections/results/performance}
\input{sections/results/generalizability}

%% file: sections/results/consistency.tex
\subsection{RQ1. Consistency}\label{sec:results-consistency}

To assess the consistency of ChatGPT in correctly screening articles, we run our experiments on the same conditions multiple times and observe key statistical dispersion metrics (\secref{sec:dispersion}) and moments, and calculate agreement metrics (\secref{sec:agreement}) of the different runs.
We report consistency w.r.t. MCC as it has been shown to be the only metric that gives a high score if all indicators perform well: high TP and TN, low FN and FN~\citep{Chicco2023}. MCC also works well on imbalanced data and is correlated with balanced accuracy.
We note that consistency w.r.t. other metrics follow the same trend as w.r.t. MCC. Detailed data is available in \appref{app:consistency}.

\subsubsection{Statistical dispersion}\label{sec:dispersion}

We use two measures of dispersion and one measure of outlier likelihood.

\begin{description}[leftmargin=1em]
    \item[Standard deviation] measures the dispersion of a dataset relative to its mean.
    \item[Interquartile range (IQR)] is another measure of dispersion, and it is defined as the difference between the 75th and 25th percentiles of the data, where the 50th percentile is the median.
    \item[Kurtosis] is a measure of the tailedness of a distribution, where tailedness means how often outliers occur.
\end{description}

For our purposes, in each case, lower values are better.

\input{tables/consistency-moments}

Tables \ref{tab:consistency-rl4se} and \ref{tab:consistency-dsmlcompo} show the consistency metrics of the classifiers w.r.t. MCC.
We observe similar patterns across the two datasets.
All traditional classifiers exhibit substantial dispersion across different runs, whereas the ChatGPT shows excellent consistency.
In fact, the mean and median in the ChatGPT experiments are almost identical, which is the artifact of negligible standard deviation and interquartile range. A kurtosis value below 0 is an indicator of no outliers in the data set---which is exactly the case in the ChatGPT experiments. Only two classifiers, ChatGPT and Complement Naive Bayes score under 0 in both cases. However, the Standard deviation of Complement Naive Bayes is an order of magnitude higher than that of ChatGPT.

\subsubsection{Inter-rater agreement}\label{sec:agreement}

To further assess consistency, we compute the Fleiss' kappa inter-rater agreement metric among the 10 runs of ChatGPT experiments. Fleiss' kappa characterizes the agreement of decisions and ranges between 0.0 and 1.0 with lower values corresponding to poor agreement and higher values to better agreement. Values in the 0.81--1.00 range are considered almost perfect agreement.

\input{tables/consistency-kappa}

As reported in Table \ref{tab:kappa}, ChatGPT scores substantially higher in Fleiss' kappa than the traditional classifiers. Both the $0.821$ value of RL4SE and the $0.973$ value of DSMLCompo are in the ``almost perfect agreement'' range.
In contrast, the kappa oscillates between 0.22 and 0.55 for the traditional classifiers, indicating only ``fair'' to ``moderate'' agreement.
The high agreement metric is another indicator of ChatGPT's consistency in screening articles in SR.

Although the agreement between the different runs of ChatGPT is substantial, it is interesting to note that even with a temperature set to 0, the runs are not in perfect agreement.
In the RL4SE dataset, there are 18\% (195) occurrences where at least one run outputs a different decision than the majority.
Whereas, it is 3\% (84) in the DSMLCompo dataset.
Most disagreements are due to false positives, only very few (2\%) are due to false negatives.
Thus, when disagreements occur, it is because ChatGPT includes an article that it should not. Therefore, the specificity fluctuates slightly between different runs, but recall remains identical.

\begin{conclusionframe}{Conclusion}
From these observations, we conclude that \textbf{ChatGPT screens the same articles consistently with the same decision}, substantially more consistently than traditional classifiers. On very few occasions, it may produce a different decision, usually to include an article conservatively.
\end{conclusionframe}

\phantom{}

Utilizing this conclusion, in the following research questions, we rely on a single run of ChatGPT to save costs incurred by using the API of ChatGPT.

%% file: tables/consistency-moments.tex
\begin{table*}[h!]
\centering
\caption{Moment statistics of the MCC scores for the RL4SE dataset (N=10). Bold is best.}
\label{tab:consistency-rl4se}
  \begin{tabular}{@{}lccccc@{}}
    \toprule
    \multicolumn{1}{c}{\textbf{Model}} &
    \multicolumn{1}{c}{\textbf{Mean}} &
    \multicolumn{1}{c}{\textbf{Median}} &
    \multicolumn{1}{c}{\textbf{Std. dev.}} &
    \multicolumn{1}{c}{\textbf{IQR}} &
    \multicolumn{1}{c}{\textbf{Kurtosis}} \\
    \midrule
    Random                        & 0.503 & 0.502 & 0.011 & 0.015 & -0.028 \\
    \midrule
    Logistic Regression           & 0.660 & 0.663 & 0.046 & 0.067 & 0.085 \\
    Complement Naive Bayes        & 0.648 & 0.649 & 0.031 & 0.053 & -0.375 \\
    Support Vector Classification & 0.649 & 0.629 & 0.048 & 0.050 & 1.723 \\
    Random Forest                 & 0.641 & 0.635 & 0.028 & 0.044 & 0.524 \\
    \midrule
    GPT-3.5                       & 0.649 & 0.648 & \textbf{0.002} & \textbf{0.005} & \textbf{-1.236} \\
    \bottomrule
\end{tabular}
\end{table*}

\begin{table*}[h!]
  \centering
  \caption{Moment statistics of the MCC scores for the DSMLCompo dataset (N=10)}
  \label{tab:consistency-dsmlcompo}
    \begin{tabular}{@{}lccccc@{}}
      \toprule
      \multicolumn{1}{c}{\textbf{Model}} &
      \multicolumn{1}{c}{\textbf{Mean}} &
      \multicolumn{1}{c}{\textbf{Median}} &
      \multicolumn{1}{c}{\textbf{Std. dev.}} &
      \multicolumn{1}{c}{\textbf{IQR}} &
      \multicolumn{1}{c}{\textbf{Kurtosis}} \\
      \midrule
      Random                        & 0.502 & 0.501 & 0.007 & 0.013 & -0.440 \\
      \midrule
      Logistic Regression           & 0.600 & 0.597 & 0.012 & 0.021 & -1.003 \\
      Complement Naive Bayes        & 0.603 & 0.603 & 0.013 & 0.024 & -1.320 \\
      Support Vector Classification & 0.598 & 0.599 & 0.010 & 0.016 & -0.686 \\
      Random Forest                 & 0.602 & 0.600 & 0.013 & 0.010 & 4.199 \\
      \midrule
      GPT-3.5                       & 0.628 & 0.627 & \textbf{0.001} & \textbf{0.003} & \textbf{-1.350} \\
      \bottomrule
    \end{tabular}
\end{table*}

%% file: tables/consistency-kappa.tex
\begin{table*}[h!]
  \centering
  \caption{Fleiss' Kappa inter-rater agreement scores over 10 runs. Bold is best.}
  \label{tab:kappa}
    \begin{tabular}{@{}lcc@{}}
      \toprule
      \multicolumn{1}{c}{\textbf{Model}} &
      \multicolumn{1}{c}{\textbf{RL4SE}} &
      \multicolumn{1}{c}{\textbf{DSMLCompo}} \\
      \midrule
      Random & -0.007 & -0.004 \\
      \midrule
      Logistic Regression & 0.223 & 0.354 \\
      Complement Naive Bayes & 0.356 & 0.550 \\
      Support Vector Classification & 0.251 & 0.301 \\
      Random Forest & 0.353 & 0.419 \\
      \midrule
      GPT-3.5        & \textbf{0.821} & \textbf{0.973} \\
      \bottomrule
    \end{tabular}
\end{table*}

%% file: sections/results/performance.tex
\subsection{RQ2. Classification performance}\label{sec:results-performance}

We report the results of the classification performance of ChatGPT in comparison to traditional classifiers for each of the five data sets.
For the two large datasets, RL4SE and DSMLCompo, we also report significance figures as we ran 10 experiments previously.
Due to RQ1, we conducted only one processed each article once with ChatGPT for the other three datasets (MobileMDE, MPM4CPS, UpdateCollabMDE).
Therefore, significance analysis is not available for these datasets.

\subsubsection{RL4SE}
\input{tables/classifiers-rl4se}
Table \ref{tab:classifiers-rl4se} reports the results for the RL4SE dataset averaged over 10 runs.

On the one hand, we note low precision and high recall scores for all classifiers. This means that all the classifiers, including ChatGPT, tend to include too many articles, but rarely exclude articles incorrectly.
This is expected since we favored recall over precision while training the classifiers (with F2) and prompt engineering for ChatGPT.
F2 values barely reach 50\% confirming the imbalance between precision and recall.
On the other hand, NPV and specificity are higher indicating that the classifiers are more accurate to exclude articles than to include them.

Traditional classifiers scored similarly on all metrics. For this dataset, we note that ChatGPT has higher scores for recall, NPV, balanced accuracy, and F2, while Logistic Regression has higher scores for precision, specificity, and MCC. All classifiers perform better than Random.

Overall, balanced accuracy is around 72\% for the four traditional classifiers and reaches 75\% with ChatGPT.
MCC scores are similar among traditional classifiers and ChatGPT.

\subsubsection{DSMLCompo}
\input{tables/classifiers-dsmlcompo}
Table \ref{tab:classifiers-dsmlcompo} reports the results for the DSMLCompo dataset averaged over 10 runs.

We observe the same trend of high recall and low precision scores for all classifiers.
Compared to the RL4SE dataset, the gap between both metrics is higher.
The values of specificity and NPV are slightly higher than for RL4SE, with almost perfect scores for NPV for each classifier. Again, this indicates that the classifiers correctly exclude articles.
For this dataset, we note that ChatGPT has higher scores for all metrics, except specificity which is higher with Random Forest.
The balanced accuracy and MCC of all classifiers is similar in both datasets.

\subsubsection{Significance analysis of classification performance on the RL4SE and DSMLCompo data sets}

A Shapiro-Wilk test indicates that most of our variables are parametric.
Thus, we perform one-way ANOVAs with post hoc tests to assess between-group mean differences at $\alpha=0.05$ using SPSS 28. Detailed data is available in \appref{app:significance}.

Both in RL4SE and DSMLCompo, ChatGPT performs significantly better than Random in each metric.
For the RL4SE dataset, our results show that, overall, there is no significant difference between ChatGPT and any of the four traditional classifiers. Nevertheless, it performs significantly better than Logistic Regression and Random Forest for recall and NPV.
For the DSMLCompo dataset, our results show that ChatGPT performs significantly better than every traditional classifier for NPV, F2, bAcc, and MCC.
Additionally, ChatGPT performs significantly better than  Logistic Regression for precision. It also performs significantly better than Random Forest and Support Vector Classification for recall.

\subsubsection{UpdateCollabMDE}
\input{tables/classifiers-updatecollabmde}

Table \ref{tab:classifiers-updatecollabmde} reports the results for the UpdateCollabMDE dataset with the traditional classifiers averaged over 10 runs and ChatGPT with 1 run (leveraging the results of RQ1).

We observe a similar performance trend as in the previous two datasets with high recall and NPV, low precision, and moderate specificity.
That is, ChatGPT includes everything that needs to be included but also includes articles that should have been excluded. Conversely, ChatGPT only excludes articles that should be excluded, but not everything that should be excluded.
ChatGPT performs poorly in precision, but this is generally true for other classifiers as well.
For this dataset, we note that ChatGPT has higher scores for recall and NPV only.
Complement Naive Bayes seems to perform best on all other metrics.

Overall, ChatGPT performs comparably to traditional classifiers, as evidenced by the minuscule difference from the best balanced accuracy, F2, and MCC numbers.
However, we note that all classifiers have much lower scores on these three metrics than for RL4SE and DSMLCompo. In particular, ChatGPT misses a lot of articles to exclude.

\subsubsection{MobileMDE}
\input{tables/classifiers-mobilemde}
Table \ref{tab:classifiers-mobilemde} reports the results for the MobileMDE dataset with the traditional classifiers averaged over 10 runs and ChatGPT with 1 run (leveraging the results of RQ1).

We observe much higher specificity by ChatGPT than by traditional classifiers. ChatGPT's specificity for this dataset is almost perfect, which is also higher than in all the other datasets. This means that it excludes almost all the articles that should be excluded. However, it performs particularly poorly in recall, with a score even lower than random. Precision is still the highest with ChatGPT, which means that when ChatGPT includes an article, this article should indeed be included. The NPV, although still high, is the lowest with ChatGPT among all classifiers.
Balanced accuracy and F2 are lowest with ChatGPT, while MCC is similar to the other classifiers.
Interestingly, all classifiers obtain worse MCC scores than random for this dataset.
Like for the UpdateCollabMDE dataset, Complement Naive Bayes seems to perform best on all other metrics.

Overall, ChatGPT performs comparably to traditional classifiers, although it misses a lot of articles to include.

\subsubsection{MPM4CPS}
\input{tables/classifiers-mpm4cps}

Table \ref{tab:classifiers-mpm4cps} reports the results for the MPM4CPS dataset with the traditional classifiers averaged over 10 runs and ChatGPT with 1 run (leveraging the results of RQ1).

In this dataset, no classifier performs better than the others. In fact, each classifier is better than the others for only one metric.
Overall, ChatGPT still performs comparably to traditional classifiers. While it has the highest scores for balanced accuracy and F2, it comes second for MCC.

\begin{conclusionframe}{Conclusion}
From these observations, we conclude that \textbf{the classification performance of ChatGPT is comparable to that of traditional classifiers}. It \textbf{rarely misses articles to include and excludes most articles} that should be excluded. In general, its classification performance to correctly include articles is above 70\% and to correctly exclude is above 60\%.
\end{conclusionframe}

%% file: tables/classifiers-rl4se.tex
\begin{table*}[h!]
  \centering
  \caption{Classifiers and their performance on the RL4SE dataset (N=10). Bold is best.}
  \label{tab:classifiers-rl4se}
  \begin{tabular}{@{}lccccccccc@{}}
    \toprule
    \multicolumn{1}{c}{\textbf{Model}} &
    \multicolumn{1}{c}{\textbf{Rec}} &
    \multicolumn{1}{c}{\textbf{Prec}} &
    \multicolumn{1}{c}{\textbf{Spec}} &
    \multicolumn{1}{c}{\textbf{NPV}} &
    \multicolumn{1}{c}{\textbf{bAcc}} &
    \multicolumn{1}{c}{\textbf{F2}} &
    \multicolumn{1}{c}{\textbf{MCC}} \\
    \midrule
    Random                        & 0.515 & 0.088 & 0.497 & 0.916 & 0.506 & 0.262 & 0.503 \\
    \midrule
    Logistic Regression           & 0.643 & \textbf{0.274} & \textbf{0.804} & 0.960 & 0.723 & 0.486 & \textbf{0.660} \\
    Complement Naive Bayes        & 0.715 & 0.228 & 0.740 & 0.966 & 0.727 & 0.483 & 0.648 \\
    Support Vector Classification & 0.736 & 0.233 & 0.717 & 0.967 & 0.726 & 0.485 & 0.649 \\
    Random Forest                 & 0.689 & 0.222 & 0.748 & 0.963 & 0.719 & 0.471 & 0.641 \\
    \midrule
    GPT-3.5              & \textbf{0.821} & 0.199 & 0.688 & \textbf{0.976} & \textbf{0.755} & \textbf{0.505} & 0.649 \\
    \bottomrule
  \end{tabular}
\end{table*}

%% file: tables/classifiers-dsmlcompo.tex
\begin{table*}[h!]
  \centering
  \caption{Classifiers and their performance on the DSMLCompo dataset (N=10). Bold is best.}
  \label{tab:classifiers-dsmlcompo}
  \begin{tabular}{@{}lccccccccc@{}}
    \toprule
    \multicolumn{1}{c}{\textbf{Model}} &
    \multicolumn{1}{c}{\textbf{Rec}} &
    \multicolumn{1}{c}{\textbf{Prec}} &
    \multicolumn{1}{c}{\textbf{Spec}} &
    \multicolumn{1}{c}{\textbf{NPV}} &
    \multicolumn{1}{c}{\textbf{bAcc}} &
    \multicolumn{1}{c}{\textbf{F2}} &
    \multicolumn{1}{c}{\textbf{MCC}} \\
    \midrule
    Random                        & 0.508 & 0.057 & 0.499 & 0.945 & 0.504 & 0.196 & 0.502 \\
    \midrule
    Logistic Regression           & 0.807 & 0.112 & 0.614 & 0.982 & 0.711 & 0.358 & 0.600 \\
    Complement Naive Bayes        & 0.811 & 0.116 & 0.619 & 0.983 & 0.715 & 0.364 & 0.603 \\
    Support Vector Classification & 0.770 & 0.114 & 0.639 & 0.979 & 0.704 & 0.356 & 0.598 \\
    Random Forest                 & 0.747 & 0.121 & \textbf{0.670} & 0.978 & 0.708 & 0.364 & 0.602 \\
    \midrule
    GPT-3.5              & \textbf{0.869} & \textbf{0.133} & 0.666 & \textbf{0.988} & \textbf{0.767} & \textbf{0.413} & \textbf{0.628} \\
    \bottomrule
  \end{tabular}
\end{table*}

%% file: tables/classifiers-updatecollabmde.tex
\begin{table*}[h!]
\centering
\caption{Classifiers and their performance on the UpdateCollabMDE dataset (N$_{\text{traditional}}$=10, N$_{\text{ChatGPT}}$=1). Bold is best.}
\label{tab:classifiers-updatecollabmde}
\begin{tabular}{@{}lccccccccc@{}}
    \toprule
    \multicolumn{1}{c}{\textbf{Model}} &
    \multicolumn{1}{c}{\textbf{Rec}} &
    \multicolumn{1}{c}{\textbf{Prec}} &
    \multicolumn{1}{c}{\textbf{Spec}} &
    \multicolumn{1}{c}{\textbf{NPV}} &
    \multicolumn{1}{c}{\textbf{bAcc}} &
    \multicolumn{1}{c}{\textbf{F2}} &
    \multicolumn{1}{c}{\textbf{MCC}} \\
    \midrule
    Random                        & 0.482 & 0.063 & 0.498 & 0.932 & 0.490 & 0.207 & 0.495 \\
    \midrule
    Logistic Regression           & 0.775 & 0.133 & 0.602 & 0.975 & 0.689 & 0.380 & 0.600 \\
    Complement Naive Bayes        & 0.704 & \textbf{0.169} & \textbf{0.705} & 0.971 & \textbf{0.704} & \textbf{0.406} & \textbf{0.617} \\
    Support Vector Classification & 0.602 & 0.098 & 0.631 & 0.965 & 0.616 & 0.279 & 0.564 \\
    Random Forest                 & 0.709 & 0.147 & 0.661 & 0.971 & 0.685 & 0.382 & 0.603 \\
    \midrule
    GPT-3.5                       & \textbf{0.947} & 0.108 & 0.455 & \textbf{0.992} & 0.701 & 0.371 & 0.600 \\
    \bottomrule
\end{tabular}
\end{table*}

%% file: tables/classifiers-mobilemde.tex
\begin{table*}[h!]
  \centering
  \caption{Classifiers and their performance on the MobileMDE dataset (N$_{\text{traditional}}$=10, N$_{\text{ChatGPT}}$=1). Bold is best.}
  \label{tab:classifiers-mobilemde}
  \begin{tabular}{@{}lccccccccc@{}}
    \toprule
    \multicolumn{1}{c}{\textbf{Model}} &
    \multicolumn{1}{c}{\textbf{Rec}} &
    \multicolumn{1}{c}{\textbf{Prec}} &
    \multicolumn{1}{c}{\textbf{Spec}} &
    \multicolumn{1}{c}{\textbf{NPV}} &
    \multicolumn{1}{c}{\textbf{bAcc}} &
    \multicolumn{1}{c}{\textbf{F2}} &
    \multicolumn{1}{c}{\textbf{MCC}} \\
    \midrule
    Random                       & 0.505 & 0.189 & 0.495 & 0.812 & 0.500 & 0.378 & 0.500 \\
    \midrule
    Logistic Regression           & 0.660 & 0.369 & 0.695 & 0.904 & 0.677 & 0.543 & 0.654 \\
    Complement Naive Bayes        & \textbf{0.682} & 0.390 & 0.733 & \textbf{0.911} & \textbf{0.708} & \textbf{0.580} & \textbf{0.676} \\
    Support Vector Classification & 0.593 & 0.292 & 0.673 & 0.884 & 0.633 & 0.473 & 0.613 \\
    Random Forest                 & 0.580 & 0.363 & 0.737 & 0.885 & 0.658 & 0.504 & 0.640 \\
    \midrule
    GPT-3.5                & 0.327 & \textbf{0.514} & \textbf{0.928} & 0.856 & 0.628 & 0.353 & 0.654 \\
    \bottomrule
  \end{tabular}
\end{table*}

%% file: tables/classifiers-mpm4cps.tex
\begin{table*}[h!]
\centering
\caption{Classifiers and their performance on the MPM4CPS dataset (N$_{\text{traditional}}$=10, N$_{\text{ChatGPT}}$=1). Bold is best.}
\label{tab:classifiers-mpm4cps}
\begin{tabular}{@{}lccccccccc@{}}
    \toprule
    \multicolumn{1}{c}{\textbf{Model}} &
    \multicolumn{1}{c}{\textbf{Rec}} &
    \multicolumn{1}{c}{\textbf{Prec}} &
    \multicolumn{1}{c}{\textbf{Spec}} &
    \multicolumn{1}{c}{\textbf{NPV}} &
    \multicolumn{1}{c}{\textbf{bAcc}} &
    \multicolumn{1}{c}{\textbf{F2}} &
    \multicolumn{1}{c}{\textbf{MCC}} \\
    \midrule
    Random                        & 0.504 & 0.527 & 0.501 & 0.478 & 0.502 & 0.508 & 0.502 \\
    \midrule
    Logistic Regression           & \textbf{0.746} & 0.643 & 0.518 & 0.662 & 0.632 & 0.714 & 0.642 \\
    Complement Naive Bayes        & 0.582 & 0.637 & \textbf{0.619} & 0.596 & 0.601 & 0.581 & 0.607 \\
    Support Vector Classification & 0.638 & 0.597 & 0.553 & 0.617 & 0.596 & 0.618 & 0.601 \\
    Random Forest                 & 0.713 & \textbf{0.689} & 0.605 & \textbf{0.684} & 0.659 & 0.694 & \textbf{0.672} \\
    \midrule
    GPT-3.5                & 0.738 & 0.664 & 0.592 & 0.674 & \textbf{0.665} & \textbf{0.722} & 0.667 \\
    \bottomrule
\end{tabular}
\end{table*}

%% file: sections/results/generalizability.tex
\subsection{RQ3. Generalizability}\label{sec:results-generalizability}
\input{tables/generalizability}

\tabref{tab:generalizability} aggregates the classification performance of ChatGPT from Tables \ref{tab:classifiers-rl4se}--\ref{tab:classifiers-mpm4cps}.
The performance profile of ChatGPT substantially varies across different datasets. For example, while recall is low in the MobileMDE dataset, it is three times higher in the UpdateCollabMDE dataset. However, it is the opposite for specificity: low for the latter and twice as high for the former.
Another example is precision, which is very low in the UpdateCollabMDE dataset, while it is six times higher in the MPM4CPS dataset.
NPV seems to have consistently high scores.
Interestingly, we observe similar variations for all classifiers across the datasets.

According to the moment statistics in \tabref{tab:generalizability}, the only generalizable metric across all five datasets is balanced accuracy with an average classification performance of 70\%.

\begin{figure*}[t]
  \centering
  \begin{subfigure}[b]{0.32\textwidth}
    \centering
    \includegraphics[trim={1.5cm 0.5cm 1.5cm 0.5cm},clip,width=\textwidth]{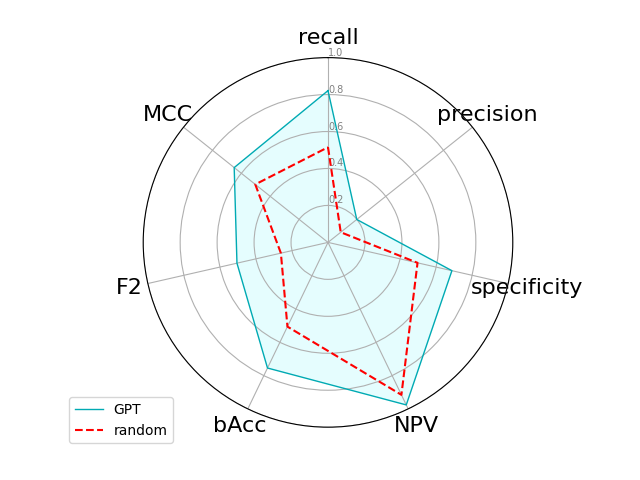}
    \caption{RL4SE}
    \label{fig:radar-rl4se}
  \end{subfigure}
  \hfill
  \begin{subfigure}[b]{0.32\textwidth}  
    \centering 
    \includegraphics[trim={1.5cm 0.5cm 1.5cm 0.5cm},clip,width=\textwidth]{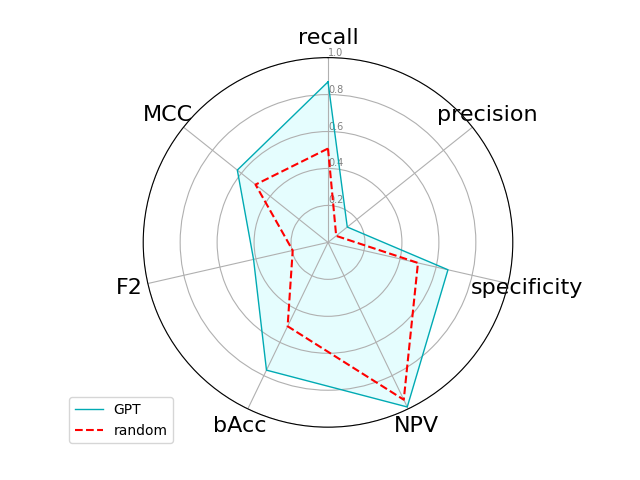}
    \caption{DSMLCompo}
    \label{fig:radar-dsmlcompo}
  \end{subfigure}
  \hfill
  \begin{subfigure}[b]{0.32\textwidth}   
    \centering 
    \includegraphics[trim={1.5cm 0.5cm 1.5cm 0.5cm},clip,width=\textwidth]{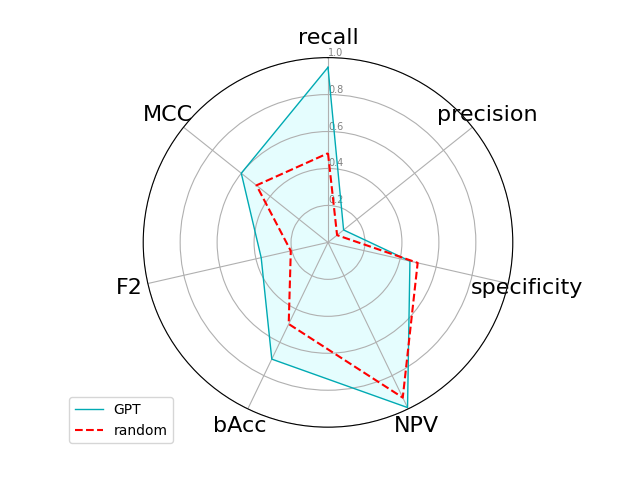}
    \caption{UpdateCollabMDE}
    \label{fig:radar-updatecollabmde}
  \end{subfigure}
  \hfill
  \begin{subfigure}[b]{0.32\textwidth}  
    \centering
  \end{subfigure}
  \vskip\baselineskip
  \begin{subfigure}[b]{0.32\textwidth}   
    \centering 
    \includegraphics[trim={1.5cm 0.5cm 1.5cm 0.5cm},clip,width=\textwidth]{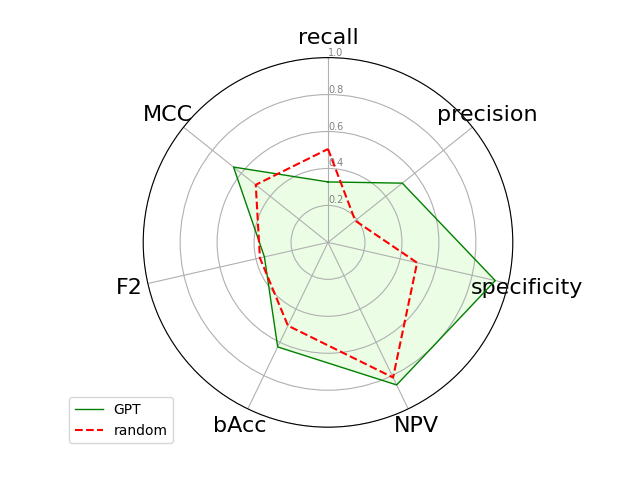}
    \caption{MobileMDE}
    \label{fig:radar-mobilemde}
  \end{subfigure}
  \qquad
  \begin{subfigure}[b]{0.32\textwidth}
    \centering 
    \includegraphics[trim={1.5cm 0.5cm 1.5cm 0.5cm},clip,width=\textwidth]{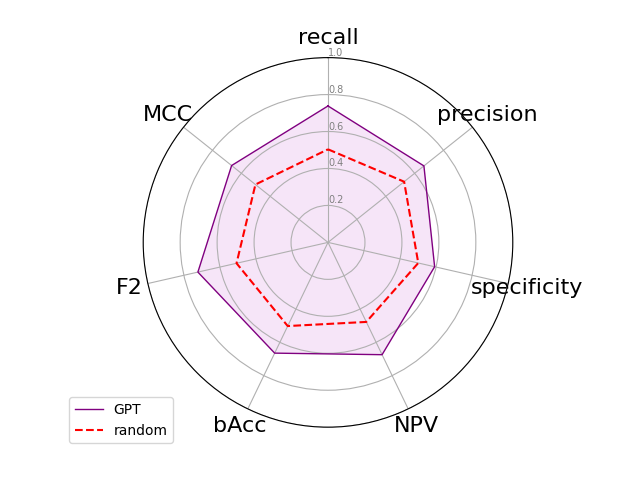}
    \caption{MPM4CPS}
    \label{fig:radar-mpm4cps}
  \end{subfigure}
  \caption{Radar plots showing the performance profile of ChatGPT on different data sets. (Color coding corresponds to \figref{fig:bubble}.)} 
  \label{fig:radar}
\end{figure*}
These performance profiles are more intuitive when visualized on radar plots, shown in \figref{fig:radar}.
The performance of ChatGPT is very similar in the RL4SE (\figref{fig:radar-rl4se}), DSMLCompo (\figref{fig:radar-dsmlcompo}), and UpdateCollabMDE (\figref{fig:radar-updatecollabmde}) datasets, with high recall and NPV, but low precision, and moderate specificity.
This performance profile is seemingly general for datasets with usual inclusion ($<10\%$) and conflict ratios ($<10\%$), as shown in \figref{fig:bubble} with the corresponding color coding.
This profile tends to be on the safe side of things: not losing articles (high recall and high NPV ensure this), but potentially including articles that should have been excluded (low precision).

A characteristically different performance profile appears for the case of MobileMDE (\figref{fig:radar-mobilemde}). Recall that this dataset has more than $50\%$ conflict and a higher inclusion rate close to $20\%$. (See the green cluster in \figref{fig:bubble}.) While ChatGPT excludes articles correctly (high specificity and NPV), it misses a significant amount of articles to include (recall worse than random)and incorrectly includes too many articles (precision around $50\%$).
This profile tends to aggressively exclude articles even if they should have been included. However, the high conflict rate among the reviewers may explain why ChatGPT performs as such.

MPM4CPS is a balanced dataset with almost the same number of articles included and excluded. (See the purple cluster in \figref{fig:bubble}.) The performance profile of ChatGPT for this dataset  (\figref{fig:radar-mpm4cps}) is rather balanced with an average score around 60\% for all metrics.
Still better than random, this performance profile provides a trade-off between safe and aggressive profiles.

\begin{conclusionframe}{Conclusion}
From these observations, we conclude that the comparable classification performance of ChatGPT \textbf{generally translates to multiple datasets}. This means, no re-training is needed like in the case of a traditional classifier, and no prompt re-engineering is needed either.
However, ChatGPT also exhibits \textbf{different performance profiles} on datasets with different inclusion and conflict ratios. Therefore, its generalization is only applicable to datasets with similar characteristics.
\end{conclusionframe}

%% file: tables/generalizability.tex
\begin{table*}[h!]
\centering
\caption{ChatGPT performance on the five datasets. Bold is best.}
\label{tab:generalizability}
\begin{tabular}{@{}lccccccccc@{}}
  \toprule
  \multicolumn{1}{c}{\textbf{Dataset}} &
  \multicolumn{1}{c}{\textbf{Rec}} &
  \multicolumn{1}{c}{\textbf{Prec}} &
  \multicolumn{1}{c}{\textbf{Spec}} &
  \multicolumn{1}{c}{\textbf{NPV}} &
  \multicolumn{1}{c}{\textbf{bAcc}} &
  \multicolumn{1}{c}{\textbf{F2}} &
  \multicolumn{1}{c}{\textbf{MCC}} \\
  \midrule
  RL4SE & 0.821 & 0.199 & 0.688 & 0.976 & 0.755 & 0.505 & 0.649 \\
  DSMLCompo & 0.869 & 0.133 & 0.666 & 0.988 & \textbf{0.767} & 0.413 & 0.628 \\
  UpdateCollabMDE & \textbf{0.947} & 0.108 & 0.455 & \textbf{0.992} & 0.701 & 0.371 & 0.600 \\
  MobileMDE & 0.327 & 0.514 & \textbf{0.928} & 0.856 & 0.628 & 0.353 & 0.654 \\
  MPM4CPS & 0.738 & \textbf{0.664} & 0.592 & 0.674 & 0.665 & \textbf{0.722} & \textbf{0.667} \\
  \midrule
  \textbf{Mean} & 0.741 & 0.324 & 0.666 & 0.897 & 0.703 & 0.473 & 0.640 \\
  \textbf{Std. dev.} & 0.243 & 0.250 & 0.173 & 0.137 & 0.059 & 0.151 & 0.026 \\
  \textbf{Median} & 0.821 & 0.199 & 0.666 & 0.976 & 0.701 & 0.413 & 0.649 \\
  \textbf{IQR} & 0.130 & 0.381 & 0.096 & 0.132 & 0.089 & 0.134 & 0.026 \\
  \textbf{Kurtosis} & 3.199 & -2.119 & 1.492 & 1.500 & -1.958 & 1.998 & 0.075 \\
  \bottomrule
\end{tabular}
\end{table*}

%% file: sections/discussion.tex
\section{Discussion}\label{sec:discussion}

We now discuss possible interpretations and implications of the results.

\subsection{Can ChatGPT be used to assist in screening articles in an SR?}

This is the underlying question of the goal of this study.
Our results confirm the hypothesis that ChatGPT can be used to assist in screening articles in an SR.
Although it is a useful tool in screening articles, its classification performance is not sufficiently accurate to automate the process---at least not with the current prompting technique.

\subsubsection*{Similar performance---without training and feature engineering}
ChatGPT performs comparably to traditional classifiers.
Moreover, it achieves this level of classification performance without training.
Training is one of the key blockers in applying traditional classifiers as a training dataset usually becomes available after a substantial amount of manual classification, basically defeating the purpose of automation.

\subsubsection*{ChatGPT's classification performance translates to other datasets}
This is a major upgrade over traditional classifiers that have to be trained on a dataset first and re-trained for other datasets. LLMs come with pre-trained models and might only require a small number of input examples to be customized for a problem (see: few-shot learning~\cite{wang2020generalizing}).

\subsubsection*{Dataset characteristics influence performance profiles}
The metrics that matter most for an SR tool, i.e., recall and specificity, are generally high in datasets with regular SR profiles (low inclusion and conflict rates like RL4SE, DSMLCompo, and UpdateCollabMDE). Balanced profiles (MPM4CPS with over 50\% inclusion ratio) exhibit a balanced classification performance with ChatGPT.
However, the accuracy is between 62\% and 77\%, which is rather low.

It is also interesting to see (in \figref{fig:radar}) that the performance profile of ChatGPT on datasets with regular SR profiles is similar to the random classifier but systemically improves on it. This similarity is also present in balanced datasets (MPM4CPS), but not in datasets with a high conflict ratio (MobileMDE).
We hypothesize that the high conflict ratio that is an artifact of decision issues of human screeners is also indicative of ChatGPT's expected issues on a particular dataset. In this dataset profile, human reviewers do not agree on a substantial number of articles to screen but eventually decide to include or exclude these articles after some discussion sessions. We note that for the MobileMDE dataset, ChatGPT did not follow the final decisions in a good portion of these articles. Using ChatGPT as a chatbot, i.e., discussing some of the conflicting articles with it, may help improve its performance.

\subsection{How much screening effort can LLM-based automation realize and how does it compare to traditional classifiers?}\label{sec:discussion-wss}

Work saved over sampling (WSS) is a frequently used non-standard metric to evaluate automated screening tools~\cite{kusa2023analysis} that balances between high recall and sufficient NPV. According to \citet{cohen2006reducing}, WSS indicates the ratio of articles that, although meet the original search criteria, reviewers do not have to read because they have been excluded by the classifier.
In particular, WSS@95 is the most common version where a fixed recall level of 95\% is achieved when 95\% of a dataset is randomly sampled, and this provides a 5\% saving for reviewers.
WSS is used to measure the reduction of human screening workload by using automation tools. The first term focuses on exclusion decisions: excluding more articles reduces the subsequent human effort. However, the second term penalizes WSS with the rate of missed articles to include.
\begin{equation}\label{eq:wss}
  \mathit{WSS@recall}=\frac{TN+FN}{TP+TN+FP+FN} - 1 + \mathit{Rec}
\end{equation}

To give an estimate of how much screening time specific classifiers save over different datasets by calculating their \textit{WSS@recall} and comparing them within one dataset. Tables \ref{tab:wss-calc-1}-- \ref{tab:wss-calc-3} show the figures we obtain.

\input{tables/wsscalc-1}
\input{tables/wsscalc-2}
\input{tables/wsscalc-3}

We report saved effort in terms of saved papers (i.e., the ones that reviewers did not have to read), and assuming a screening time of each article around 1 minute, the saved time in hours.

As the tables show, ChatGPT saves the most effort in two of the four cases (RL4SE and MobileMDE), coming in as a close second in another two cases (UpdateCollabMDE and DSMLCompo).
In the DSMLCompo dataset, ChatGPT is able to save over 24 hours of screening time, that is, three working days' worth of full-time equivalent (FTE).
The biggest advantage of ChatGPT over traditional classifiers is observed in the RL4SE dataset, in which ChatGPT saves over 50\% more effort than the second-best classifier.

As evidenced by the Random classifier's performance being close to 50\%, one must take this metric with a grain of salt. While Randomly classifying articles might save 50\% of the work, it is also very likely to produce unusable corpora. Thus, one must also keep an eye on the classification performance of a classifier when trying to justify WSS. The metric, nonetheless is indicative of the capabilities of state-of-the-art screening automation in SR.

\subsection{Costs and benefits of using ChatGPT}

Based on the token consumption we measured during our experiments and the WSS numbers above, we can calculate the approximate monetary benefits of using ChatGPT.

\input{tables/tokens}

\tabref{tab:token-stats} reports the costs and savings of using ChatGPT. In each dataset, a little over 300 tokens are consumed for screening a paper. Based on the saved papers in the WSS calculations (Tables \ref{tab:wss-calc-1}-- \ref{tab:wss-calc-3}), we calculate the papers that were not saved to obtain the number of sum tokens. At the time of writing this report, ChatGPT-3.5 is priced at USD~0.002/1k tokens. Based on this figure, we calculate the final monetary costs in USD. As demonstrated, each dataset consumes less than a dollar, some even below a dime. Finally, we calculate the savings in terms of full-time equivalent days (FTE -- 8 hours a day) based on the saved hours in Tables \ref{tab:wss-calc-1}-- \ref{tab:wss-calc-3}. The direct comparison of monetary savings is left to the reader. This can be achieved by simply multiplying FTE days with a figure representative of the reader's context. In our context, we observed that automation by ChatGPT can save about 5--6 orders more than the costs ChatGPT's usage incurs.

Our calculations do not include development and experimentation costs, which were substantial in our case but are not directly charged to end-users in an SR automation tool. Of course, the price of using ChatGPT might differ depending on the type of service (hosted or on-prem), the pricing model of the SR tool (disaggregate pricing, risk sharing, etc), and various other factors. The price of ChatGPT tokens is also increasing by an order of magnitude in newer versions and more elaborate models. However, SR automation might not need the latest and too complex models. Thus, we anticipate that the return on investment of using ChatGPT to automate SR will remain high.

\subsection{Metrics usefulness}

The literature proposes various metrics (see \secref{sec:metrics}) to evaluate binary classifiers and our results confirm that no single metric is sufficient to assess the classification performance of automated article screening in SRs.
Ultimately what counts is the number of TP, TN, FP, and FN cases. However, it cannot be used across different datasets.
MCC seems to be the most reliable metric to compare the classification performance of different classifiers within the same dataset. It is the only metric that is truly influenced by the above-mentioned four quantities.
As shown in Tables \ref{tab:classifiers-rl4se}--\ref{tab:classifiers-mpm4cps}, MCC can be useful to rank the best and worst performing classifiers.

Balanced accuracy gives a reliable estimate of the accuracy of the classifier, given that the inclusion/exclusion ratio is typically unbalanced.
When the corpus is more balanced (like the MPM4CPS dataset), it has the same value as the typical \textit{accuracy} metric.
Therefore, balanced accuracy is useful to quantify the performance of a classifier.
As shown in Table \ref{tab:generalizability}, it can be used to compare the performance of the same classifier across different datasets.

In retrospect, the F-score metric, like F2, is not really useful for this problem since it looks at the performance of including articles only.
We recommend using MCC to train and fine-tune classifiers instead.

\begin{figure}[h!]
    \centering
    \includegraphics[width=0.75\linewidth]{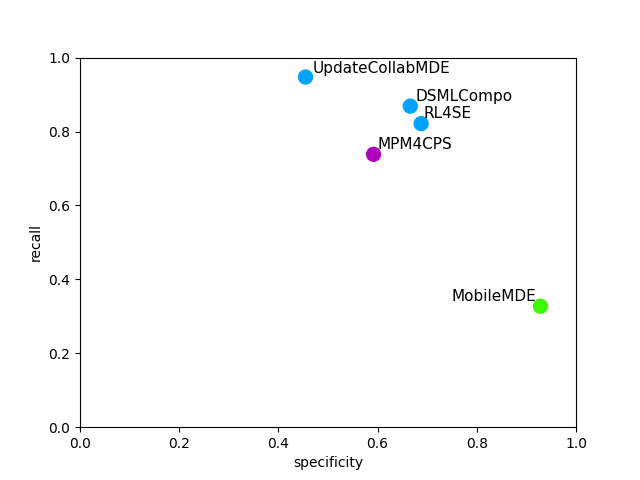}
    \caption{Recall and precision of ChatGPT, conflict, and size characteristic of the data sets. (Color coding corresponds to \figref{fig:bubble}.)}
    \label{fig:scatter}
\end{figure}

The different performance profiles are characterized mostly by the four metrics: precision, recall, NPV, and specificity.
These metrics give the most incite on the advantages and disadvantages of a classifier for a given dataset.
However, in unbalanced datasets with high exclusions and low inclusions, precision tends to always be low and NPV high.
Therefore, specificity and recall are truly the two main metrics to analyze when understanding the details of the classification performance of a classifier.
\figref{fig:scatter} shows how specificity and recall align in the five datasets. ChatGPT with the developed prompt tends to strike a balance between specificity and recall, as demonstrated by the cluster formed by the four datasets with typical SR corpus characteristics (see \figref{fig:bubble} with corresponding color coding). The unusually high conflict ratio in MobileMDE likely causes conflicting views between the human-produced ground truth and ChatGPT's decisions, resulting in subpar recall. We note that after manually reviewing some papers of the MobileMDE dataset, we were also unable to classify the sample correctly. This is likely due to the atypically broad scope of the MobileMDE project that might require more interactions among researchers to decide about an inclusion.

Finally, WSS can be used to estimate the effort saved using an automated classifier vs. manually screening articles. But it should not be used to compare classifiers as it has no minimum and maximum values \cite{kusa2023analysis}.

In conclusion, we recommend reporting recall, specificity, precision, and NPV to analyze the classification performance with a special focus on the former two.
We recommend reporting the balanced accuracy to quantify this performance.
We also recommend training the classifiers using MCC and reporting its score to compare the classification performance for different dataset profiles.

\subsection{Implications}
We foresee generative AI services, such as ChatGPT disrupting the ways secondary and tertiary studies are conducted. Especially due to its relatively low costs and high return on investment, generative AI supplied with LLMs is set to become a key element of tools literature reviews. We anticipate community standards to embrace this change and incorporate LLM-based automation into the toolbox of evidence-based software engineering research. 

Among the most important future trends, we expect the proliferation of rapid SRs that trade completeness for substantially reduced completion time~\cite{ralph2022paving}. This is especially justified in time-sensitive situations (e.g., in the preparatory phase of research projects), SRs are not feasible to be conducted.
Typical shortcuts in rapid reviews are related to the size of the corpus, such as restricting literature search, omitting snowballing, and streamlining screening~\cite{ganann2010expediting}.
With the support of LLMs, rapid reviews~\cite{ralph2022paving} can be conducted without quality compromise. Screening can be fully automated and the effort that would have been spent on screening can be used for validation, quality assessment, and fighting publication bias by snowballing.

We foresee small-team and solo SRs becoming more accepted. The prevalent community standards demand at least two reviewers to screen each article in an SR to mitigate obvious biases and threats to validity, with a third person acting as a tie-breaker if the other two cannot agree~\cite{kitchenham2004evidence,kitchenham2007guidelines}. A recent bibliometric analysis from \citet{fiala2017computer} reveals the average number of authors in software engineering articles is 2.67 with over half of the articles having one or two authors. Certainly, gathering a team for an SR is a challenge for the majority of software engineering researchers.
With LLMs such as ChatGPT being able to act as a teammate---albeit one whose work needs to be monitored and critically invigilated---SR teams of smaller size or even of size one (solo) become possible.

However, we warn that using ChatGPT to automate an SR process requires careful consideration of the emerging classification performance profile of ChatGPT on the given SR's corpus. Thus, SR tools that integrate ChatGPT must account for these specificities and proactively generate the most appropriate SR process for the corpus that researchers can follow.
This can be achieved by conducting a sufficiently voluminous pilot that allows assessing the expected inclusion and conflict ratios; thus, this helps identify the classification performance profile of ChatGPT for the given corpus.
Corpora that exhibit a low recall in the pilot phase, might require an appropriately dimensioned validation phase on the excluded articles. Conversely, high recall and low precision might necessitate validation of the included articles.
Focusing human effort on articles that ChatGPT is not sure about might further improve the soundness of the SR. 
Finally, we recommend tool builders develop functionality that allows researchers to express their desired SR strategy in terms of customizing a common prompt template, such as the one shown in \lstref{lst:prompt-template}. Based on high-level descriptions, the SR process can be generated and supported by the appropriate GUI, e.g., via web forms~\cite{bigendako2018modeling}. 

%% file: tables/wsscalc-1.tex
\begin{table}[h!]
\centering
\caption{Effort savings. Bold is best.\vspace{-10px}}
\label{tab:wss-calc-1}
\resizebox{\columnwidth}{!}{%
\begin{tabular}{@{}lrrrrrr@{}}
\toprule
 & \multicolumn{3}{c}{RL4SE}
 & \multicolumn{3}{c}{DSMLCompo} \\
 \cmidrule(r){2-4}
 \cmidrule(r){5-7}
Total number of screened papers$^{\ast}$ & \multicolumn{3}{r}{1089} & \multicolumn{3}{r}{2683} \\
Time needed for screening & \multicolumn{3}{r}{18.2 h} & 
 \multicolumn{3}{r}{44.7 h} \\
 \cmidrule(r){2-4}
 \cmidrule(r){5-7}
 &
 \multicolumn{1}{c}{WSS} &
 \multicolumn{1}{c}{Saved papers} &
 \multicolumn{1}{c}{Saved time} &
 \multicolumn{1}{c}{WSS} &
 \multicolumn{1}{c}{Saved papers} &
 \multicolumn{1}{c}{Saved time} \\
 \cmidrule(r){2-4}
 \cmidrule(r){5-7}
Random & 0.010 & 12 & 0.2 & 0.499 & 1338 & 22.3 \\
\cmidrule(r){2-4}
\cmidrule(r){5-7}
Logistic Regression & 0.408 & 444 & 7.4 & 0.590 & $1\,582$ & 26.4 \\
Complement Naive Bayes & 0.416 & 453 & 7.5 & 0.595 & $1\,596$ & 26.6 \\
Support Vector Classification & 0.414 & 450 & 7.5 & 0.616 & $1\,652$ & 27.5 \\
Random Forest & 0.399 & 434 & 7.2 & \textbf{0.646} & \textbf{$1\,733$} & \textbf{28.9} \\
\cmidrule(r){2-4}
\cmidrule(r){5-7}
GPT-3.5 & \textbf{0.644} & \textbf{701} & \textbf{11.7} & 0.636 & $1\,451$ & 24.2 \\
\bottomrule
\end{tabular}%
}
\vspace{1pt}
\caption*{$^{\ast}$Only counting papers with abstracts recorded in the dataset.}
\vspace{-15px}
\end{table}

%% file: tables/wsscalc-2.tex
\begin{table}[h!]
\centering
\caption{Effort savings (continued from \tabref{tab:wss-calc-1})\vspace{-10px}}
\label{tab:wss-calc-2}
\begin{tabular}{@{}lrrrrrr@{}}
\toprule
 & \multicolumn{3}{c}{UpdateCollabMDE} \\
 \cmidrule(r){2-4}
Total number of screened papers$^{\ast}$ & \multicolumn{3}{r}{875} \\
Time needed for screening & \multicolumn{3}{r}{14.6 h} \\
 \cmidrule(r){2-4}
 &
 \multicolumn{1}{c}{WSS} &
 \multicolumn{1}{c}{Saved papers} &
 \multicolumn{1}{c}{Saved time} \\
 \cmidrule(r){2-4}
Random & -0.019 & N/A & 2.4 \\
\cmidrule(r){2-4}
Logistic Regression & 0.353 & 308 & 5.1 \\
Complement Naive Bayes & \textbf{0.382} & \textbf{334} & \textbf{5.6} \\
Support Vector Classification & 0.118 & 103 & 1.7 \\
Random Forest & 0.346 & 302 & 5.0 \\
\cmidrule(r){2-4}
GPT-3.5 & 0.376 & 329 & 5.5 \\
\bottomrule
\end{tabular}%
\vspace{1pt}
\caption*{$^{\ast}$Only counting papers with abstracts recorded in the dataset.}
\vspace{-10px}
\end{table}

%% file: tables/wsscalc-3.tex
\begin{table}[h!]
\centering
\caption{Effort savings (continued from \tabref{tab:wss-calc-2})\vspace{-10px}}
\label{tab:wss-calc-3}
\resizebox{\columnwidth}{!}{%
\begin{tabular}{@{}lrrrrrr@{}}
\toprule
 & \multicolumn{3}{c}{MobileMDE}
 & \multicolumn{3}{c}{MPM4CPS} \\
 \cmidrule(r){2-4}
 \cmidrule(r){5-7}
Total number of screened papers$^{\ast}$ & \multicolumn{3}{r}{292} & \multicolumn{3}{r}{205} \\
Time needed for screening & \multicolumn{3}{r}{4.9 h} & 
 \multicolumn{3}{r}{3.4 h} \\
 \cmidrule(r){2-4}
 \cmidrule(r){5-7}
 &
 \multicolumn{1}{c}{WSS} &
 \multicolumn{1}{c}{Saved papers} &
 \multicolumn{1}{c}{Saved time} &
 \multicolumn{1}{c}{WSS} &
 \multicolumn{1}{c}{Saved papers} &
 \multicolumn{1}{c}{Saved time} \\
 \cmidrule(r){2-4}
 \cmidrule(r){5-7}
Random & 0.495 & 144 & 2.4 & 0.499 & 102 & 1.7 \\
\cmidrule(r){2-4}
\cmidrule(r){5-7}
Logistic Regression & 0.628 & 183 & 3.1 & 0.380 & 77 & 1.3 \\
Complement Naive Bayes & 0.655 & 191 & 3.2 & \textbf{0.514} & \textbf{105} & \textbf{1.8} \\
Support Vector Classification & 0.623 & 181 & 3.0 & 0.453 & 92 & 1.5 \\
Random Forest & 0.677 & 197 & 3.3 & 0.439 & 89 & 1.5 \\
\cmidrule(r){2-4}
\cmidrule(r){5-7}
GPT-3.5 & \textbf{0.880} & \textbf{256} & \textbf{4.3} & 0.419 & 85 & 1.4 \\
\bottomrule
\end{tabular}%
}
\vspace{1pt}
\caption*{$^{\ast}$Only counting papers with abstracts recorded in the dataset.}
\end{table}

%% file: tables/tokens.tex
\begin{table*}[h!]
\centering
\caption{Token consumption statistics of ChatGPT in our experiments in the different datasets}
\label{tab:token-stats}
\begin{tabular}{@{}lcccccc@{}}
  \toprule
  \multicolumn{1}{c}{\multirow{2}{*}{\textbf{Dataset}}}
  & \multicolumn{4}{c}{\textbf{Costs}}
  & \multicolumn{2}{c}{\textbf{Savings}} \\
  \cmidrule(r){2-5}
  \cmidrule(r){6-7}
  & \multicolumn{1}{c}{\textbf{Mean tokens}}
  & \multicolumn{1}{c}{\textbf{Papers not saved with WSS}}
  & \multicolumn{1}{c}{\textbf{Sum tokens}}
  & \multicolumn{1}{c}{\textbf{USD}}
  & \multicolumn{1}{c}{\textbf{Hours}}
  & \multicolumn{1}{c}{\textbf{FTE days}}\\
  \cmidrule(r){1-1}
  \cmidrule(r){2-5}
  \cmidrule(r){6-7}
  RL4SE & 343.728 & 388 & $133\,367$ & 0.267 & 18.2 & 2.275 \\
  DSMLCompo & 314.371 & $1\,232$ & $387\,305$ & 0.775 & 44.7 & 5.588 \\
  UpdateCollabMDE & 330.685 & 546 & $180\,554$ & 0.361 & 14.6 & 1.825 \\
  MobileMDE & 348.329 & 36 & $12\,540$ & 0.025 & 4.9 & 0.613 \\
  MPM4CPS & 325.932 & 120 & $39\,112$ & 0.078 & 3.4 & 0.425 \\
  \cmidrule(r){1-1}
  \cmidrule(r){2-5}
  \cmidrule(r){6-7}
  \textit{On average}& \textit{323.229} & \textit{464.4} & {$\textit{150\,108}$} & \textit{0.300} & \textit{17.16} & \textit{2.145} \\
  \bottomrule
\end{tabular}
\end{table*}

%% file: sections/conclusion.tex
\section{Conclusion}\label{sec:conclusion}

This work provides the first look at the opportunities of using ChatGPT and similar LLM for the automation of article screening in SRs. Through detailed and systematic experiments, we show that ChatGPT performs comparably in making decisions about the inclusion of articles into an SR compared to traditional classifiers.

Our results indicate that ChatGPT is a viable option to automate screening and its costs are minimal at the time of writing. Due to these beneficial qualities, we foresee a rapid adoption curve of LLMs into survey tools and novel surveying techniques to appear, e.g., solo reviewing aided by ChatGPT.

As future work, we plan to compare ChatGPT with other LLM (e.g., Alpaca\footnote{\url{https://crfm.stanford.edu/2023/03/13/alpaca.html}}$^,$\footnote{\url{https://github.com/tloen/alpaca-lora}} and Dolly\footnote{\url{https://www.databricks.com/blog/2023/03/24/hello-dolly-democratizing-magic-chatgpt-open-models.html}}), evaluate ChatGPT on a broader set of corpora, investigate further prompting techniques, and integrate an LLM into the ReLiS tool. We plan to share our findings with the creators of the used datasets and solicit feedback, potentially in the form of an interview study.

%% file: sections/acknowledgement.tex
\section*{Acknowledgement}

The authors would like to extend their gratitude to the creators of the data sets who have agreed to let us use the data they have produced in their ReLiS projects. In particular, we would like to thank
Adil Anwar,
Oussama Ben Sghaier,
Mouna Dhaouadi,
Naima Essaidi,
Jessie Galasso,
Ujjwal Hendwe,
Sebastien Mosser,
Bentley Oakes, and
Martin Weyssow
who contributed to the screening of a large body of papers in ReLiS projects that have not been published yet but served as the data to compare classifiers.

%% file: sections/appendix.tex
\appendix

\onecolumn

\input{sections/appendices/detailed-datasets}
\clearpage

\input{sections/appendices/hyperparameters}
\clearpage

\input{sections/appendices/consistency-full}

\clearpage

\input{sections/appendices/significance}

%% file: sections/appendices/detailed-datasets.tex
\section{Topic descriptions of the datasets}\label{app:datasets}

\begin{table}[!htbp]
  \caption{Data sets with topic descriptions}
  \label{tab:datasets-extended}
  {\footnotesize
  \begin{tabular}{llp{0.25\textwidth}p{0.4\textwidth}}
    \toprule
    \textbf{Project} & \textbf{Publication} & \textbf{Title} & \textbf{Topic} \\
    \toprule
    DSMLCompo       & \textit{In progress} & Domain-specific modeling language composition & approaches and techniques for composing heterogeneous domain-specific modeling languages \\
    MobileMDE       & \citet{brunschwig2022modelling} & Modeling on mobile devices & model-driven engineering techniques, languages, and tools that are touch-enabled to model software  on mobile devices \\
    MPM4CPS         & \citet{barisic2022multi} & Multi-paradigm modeling of CPS & multi-paradigm modeling approaches and applications to model cyber-physical systems \\
    RL4SE           & \textit{In progress} & Reinforcement learning for software engineering & reinforcement learning for software engineering \\
    UpdateCollabMDE & \citet{david2021collaborative} & Collaborative modeling & techniques where multiple stakeholders collaborate and manage on shared models in model-driven software engineering \\
    \bottomrule
  \end{tabular}
  }
\end{table}

%% file: sections/appendices/hyperparameters.tex
\section{Hyperparameters of the classifiers}\label{app:hyperparameters}

\subsection{General settings}

\begin{table}[h!]
\centering
\small
\caption{Hyperparameters of the classifiers}
\label{tab:tab:hyperparameters}
\begin{tabular}{@{}lll@{}}
\toprule
\multirow{2}{*}{Model} & \multicolumn{2}{l}{Parameters} \\  
 & Group & Value \\ \cmidrule(r){1-3}
\multirow{3}{*}{Logistic regression}
    & \multirow{2}{*}{grid}
        & "penalty": [None, "l2", "elasticnet"], \\
        &  & "C": numpy.logspace(-3, 3, 7), \\  \cmidrule(l){2-3} 
    & solver
        & "solver": ["lbfgs", "liblinear", "newton-cg", "newton-cholesky", "sag", "saga"] \\ \cmidrule(r){1-3} 
\multirow{1}{*}{Complement Naive Bayes}
    & \multirow{1}{*}{grid}
        & "alpha": numpy.logspace(-3, 3, 7), \\ \cmidrule(r){1-3}
\multirow{3}{*}{Support Vector Classification}
    & \multirow{3}{*}{grid}
        & "kernel": ["linear", "poly", "rbf", "sigmoid"], \\
        & & "C": numpy.logspace(-3, 3, 7), \\
        & & "gamma": ["auto", "scale"] + numpy.logspace(-3, 3, 7).tolist(), \\
        \cmidrule(r){1-3}
\multirow{5}{*}{Random Forest}
    & \multirow{5}{*}{grid}
        & "max\_depth": [80, 90, 100, 110], \\
        & & "max\_features": [2, 3], \\
        & & "min\_samples\_leaf": [3, 4, 5], \\
        & & "min\_samples\_split": [8, 10, 12], \\
        & & "n\_estimators": [100, 200, 300, 1000], \\ \bottomrule
\end{tabular}
\end{table}

For the full list of parameters (left at default), see the official documentation of the classifiers:
\begin{itemize}
    \item \url{https://scikit-learn.org/stable/modules/generated/sklearn.linear\_model.LogisticRegression.html}
    \item \url{https://scikit-learn.org/stable/modules/generated/sklearn.naive_bayes.ComplementNB.html}
    \item \url{https://scikit-learn.org/stable/modules/generated/sklearn.svm.SVC.html}
    \item \url{https://scikit-learn.org/stable/modules/generated/sklearn.ensemble.RandomForestClassifier.html}
\end{itemize}

\subsection{Settings at the specific experimental runs}

\input{tables/hyperparams}

%% file: tables/hyperparams.tex
\begin{longtblr}[caption = {Settings at the specific experimental runs}, label = {tab:hyperparams-runs},]{X[0.6,l]X[1.75,l]X[0.6,l]X[8,l]}
\hline
\textbf{Model} & \textbf{Dataset} & \textbf{Run\#} & \textbf{Parameters}  \\
    \hline
CNB & DSMLCompo & 0 & ComplementNB(alpha=10.0) \\
CNB & DSMLCompo & 1 & ComplementNB() \\
CNB & DSMLCompo & 2 & ComplementNB() \\
CNB & DSMLCompo & 3 & ComplementNB(alpha=0.001) \\
CNB & DSMLCompo & 4 & ComplementNB() \\
CNB & DSMLCompo & 5 & ComplementNB(alpha=0.1) \\
CNB & DSMLCompo & 6 & ComplementNB(alpha=0.001) \\
CNB & DSMLCompo & 7 & ComplementNB() \\
CNB & DSMLCompo & 8 & ComplementNB() \\
CNB & DSMLCompo & 9 & ComplementNB(alpha=0.001) \\
CNB & MobileMDE & 0 & ComplementNB(alpha=10.0) \\
CNB & MobileMDE & 1 & ComplementNB() \\
CNB & MobileMDE & 2 & ComplementNB() \\
CNB & MobileMDE & 3 & ComplementNB(alpha=10.0) \\ 
CNB & MobileMDE & 4 & ComplementNB(alpha=10.0) \\ 
CNB & MobileMDE & 5 & ComplementNB() \\
CNB & MobileMDE & 6 & ComplementNB() \\
CNB & MobileMDE & 7 & ComplementNB(alpha=10.0) \\ 
CNB & MobileMDE & 8 & ComplementNB(alpha=0.001) \\
CNB & MobileMDE & 9 & ComplementNB() \\
CNB & MPM4CPS & 0 & ComplementNB() \\
CNB & MPM4CPS & 1 & ComplementNB(alpha=10.0) \\
CNB & MPM4CPS & 2 & ComplementNB(alpha=10.0) \\
CNB & MPM4CPS & 3 & ComplementNB(alpha=10.0) \\
CNB & MPM4CPS & 4 & ComplementNB(alpha=10.0) \\
CNB & MPM4CPS & 5 & ComplementNB() \\
CNB & MPM4CPS & 6 & ComplementNB(alpha=10.0) \\
CNB & MPM4CPS & 7 & ComplementNB(alpha=10.0) \\
CNB & MPM4CPS & 8 & ComplementNB(alpha=10.0) \\
CNB & MPM4CPS & 9 & ComplementNB(alpha=10.0) \\
CNB & RL4SE &   0 & ComplementNB() \\
CNB & RL4SE &   1 & ComplementNB(alpha=0.001) \\
CNB & RL4SE &   2 & ComplementNB(alpha=0.001) \\
CNB & RL4SE &   3 & ComplementNB(alpha=0.001) \\
CNB & RL4SE &   4 & ComplementNB(alpha=0.001) \\
CNB & RL4SE &   5 & ComplementNB(alpha=10.0) \\
CNB & RL4SE &   6 & ComplementNB(alpha=10.0) \\
CNB & RL4SE &   7 & ComplementNB(alpha=10.0) \\
CNB & RL4SE &   8 & ComplementNB(alpha=0.001) \\
CNB & RL4SE &   9 & ComplementNB(alpha=10.0) \\
CNB & UpdateCollabMDE & 0 & ComplementNB(alpha=0.001) \\
CNB & UpdateCollabMDE & 1 & ComplementNB() \\
CNB & UpdateCollabMDE & 2 & ComplementNB(alpha=100.0) \\
CNB & UpdateCollabMDE & 3 & ComplementNB() \\
CNB & UpdateCollabMDE & 4 & ComplementNB() \\
CNB & UpdateCollabMDE & 5 & ComplementNB(alpha=0.001) \\
CNB & UpdateCollabMDE & 6 & ComplementNB(alpha=0.001) \\
CNB & UpdateCollabMDE & 7 & ComplementNB(alpha=0.001) \\
CNB & UpdateCollabMDE & 8 & ComplementNB(alpha=0.001) \\
CNB & UpdateCollabMDE & 9 & ComplementNB() \\
LR & DSMLCompo & 0 & LogisticRegression(C=1000.0, solver='liblinear') \\
LR & DSMLCompo & 1 & LogisticRegression(C=0.001, penalty=None) \\
LR & DSMLCompo & 2 & LogisticRegression(C=1000.0, solver='liblinear') \\
LR & DSMLCompo & 3 & LogisticRegression(C=1000.0, solver='liblinear') \\
LR & DSMLCompo & 4 & LogisticRegression(C=1000.0, solver='newton-cg') \\
LR & DSMLCompo & 5 & LogisticRegression(C=1000.0, solver='newton-cg') \\
LR & DSMLCompo & 6 & LogisticRegression(C=1000.0) \\
LR & DSMLCompo & 7 & LogisticRegression(C=0.001, penalty=None, solver='newton-cg') \\
LR & DSMLCompo & 8 & LogisticRegression(C=1000.0) \\
LR & DSMLCompo & 9 & LogisticRegression(C=0.001, penalty=None, solver='newton-cg') \\
LR & MobileMDE & 0 & LogisticRegression(C=0.001, penalty=None) \\
LR & MobileMDE & 1 & LogisticRegression(C=0.001, penalty=None) \\
LR & MobileMDE & 2 & LogisticRegression(C=0.001, penalty=None, solver='newton-cg') \\
LR & MobileMDE & 3 & LogisticRegression(C=0.001, penalty=None, solver='sag') \\
LR & MobileMDE & 4 & LogisticRegression(C=0.001, penalty=None, solver='newton-cg') \\
LR & MobileMDE & 5 & LogisticRegression(C=0.001, penalty=None) \\
LR & MobileMDE & 6 & LogisticRegression(C=0.001, penalty=None, solver='newton-cg') \\
LR & MobileMDE & 7 & LogisticRegression(C=0.001, penalty=None) \\
LR & MobileMDE & 8 & LogisticRegression(C=0.001, penalty=None) \\
LR & MobileMDE & 9 & LogisticRegression(C=0.001, penalty=None) \\
LR & MPM4CPS & 0 & LogisticRegression(C=1000.0, solver='saga') \\
LR & MPM4CPS & 1 & LogisticRegression(C=1000.0, solver='newton-cg') \\
LR & MPM4CPS & 2 & LogisticRegression(C=0.001, penalty=None, solver='saga') \\
LR & MPM4CPS & 3 & LogisticRegression(C=1000.0, penalty=None, solver='sag') \\
LR & MPM4CPS & 4 & LogisticRegression(C=100.0) \\
LR & MPM4CPS & 5 & LogisticRegression(C=0.001, penalty=None) \\
LR & MPM4CPS & 6 & LogisticRegression(C=100.0) \\
LR & MPM4CPS & 7 & LogisticRegression(C=0.01, penalty=None, solver='saga') \\
LR & MPM4CPS & 8 & LogisticRegression(C=1000.0, solver='liblinear') \\
LR & MPM4CPS & 9 & LogisticRegression(C=0.001, penalty=None) \\
LR & RL4SE &   0 & LogisticRegression(C=0.001, penalty=None, solver='newton-cholesky') \\
LR & RL4SE &   1 & LogisticRegression(C=0.001, penalty=None) \\
LR & RL4SE &   2 & LogisticRegression(C=1000.0, solver='newton-cg') \\
LR & RL4SE &   3 & LogisticRegression(C=0.001, penalty=None) \\
LR & RL4SE &   4 & LogisticRegression(C=0.001, penalty=None) \\
LR & RL4SE &   5 & LogisticRegression(C=0.001, penalty=None, solver='newton-cholesky') \\
LR & RL4SE &   6 & LogisticRegression(C=0.001, penalty=None, solver='newton-cholesky') \\
LR & RL4SE &   7 & LogisticRegression(C=0.001, penalty=None, solver='newton-cholesky') \\
LR & RL4SE &   8 & LogisticRegression(C=100.0) \\
LR & RL4SE &   9 & LogisticRegression(C=0.001, penalty=None, solver='newton-cholesky') \\
LR & UpdateCollabMDE & 0 & LogisticRegression(C=100.0, solver='newton-cg') \\
LR & UpdateCollabMDE & 1 & LogisticRegression(C=0.001, penalty=None) \\
LR & UpdateCollabMDE & 2 & LogisticRegression(C=0.001, penalty=None, solver='newton-cholesky') \\
LR & UpdateCollabMDE & 3 & LogisticRegression(C=1000.0) \\
LR & UpdateCollabMDE & 4 & LogisticRegression(C=0.001, penalty=None) \\
LR & UpdateCollabMDE & 5 & LogisticRegression(C=0.001, penalty=None) \\
LR & UpdateCollabMDE & 6 & LogisticRegression(C=0.001, penalty=None) \\
LR & UpdateCollabMDE & 7 & LogisticRegression(C=0.001, penalty=None, solver='newton-cg') \\
LR & UpdateCollabMDE & 8 & LogisticRegression(C=0.01, penalty=None, solver='sag') \\
LR & UpdateCollabMDE & 9 & LogisticRegression(C=0.001, penalty=None, solver='newton-cg') \\
RF & DSMLCompo & 0 & RandomForestClassifier(max\_depth=80, max\_features=2, min\_samples\_leaf=3, min\_samples\_split=8) \\
RF & DSMLCompo & 1 & RandomForestClassifier(max\_depth=80, max\_features=2, min\_samples\_leaf=3, min\_samples\_split=8) \\
RF & DSMLCompo & 2 & RandomForestClassifier(max\_depth=80, max\_features=2, min\_samples\_leaf=3, min\_samples\_split=8) \\
RF & DSMLCompo & 3 & RandomForestClassifier(max\_depth=80, max\_features=2, min\_samples\_leaf=3, min\_samples\_split=8) \\
RF & DSMLCompo & 4 & RandomForestClassifier(max\_depth=80, max\_features=2, min\_samples\_leaf=3, min\_samples\_split=8) \\
RF & DSMLCompo & 5 & RandomForestClassifier(max\_depth=80, max\_features=2, min\_samples\_leaf=3, min\_samples\_split=8) \\
RF & DSMLCompo & 6 & RandomForestClassifier(max\_depth=80, max\_features=2, min\_samples\_leaf=3, min\_samples\_split=8) \\
RF & DSMLCompo & 7 & RandomForestClassifier(max\_depth=80, max\_features=2, min\_samples\_leaf=3, min\_samples\_split=8) \\
RF & DSMLCompo & 8 & RandomForestClassifier(max\_depth=80, max\_features=2, min\_samples\_leaf=3, min\_samples\_split=8) \\
RF & DSMLCompo & 9 & RandomForestClassifier(max\_depth=80, max\_features=2, min\_samples\_leaf=3, min\_samples\_split=8) \\
RF & MobileMDE & 0 & RandomForestClassifier(max\_depth=80, max\_features=2, min\_samples\_leaf=3, min\_samples\_split=8) \\
RF & MobileMDE & 1 & RandomForestClassifier(max\_depth=80, max\_features=2, min\_samples\_leaf=3, min\_samples\_split=8) \\
RF & MobileMDE & 2 & RandomForestClassifier(max\_depth=80, max\_features=2, min\_samples\_leaf=3, min\_samples\_split=8) \\
RF & MobileMDE & 3 & RandomForestClassifier(max\_depth=80, max\_features=2, min\_samples\_leaf=3, min\_samples\_split=8) \\
RF & MobileMDE & 4 & RandomForestClassifier(max\_depth=80, max\_features=2, min\_samples\_leaf=3, min\_samples\_split=8) \\
RF & MobileMDE & 5 & RandomForestClassifier(max\_depth=80, max\_features=2, min\_samples\_leaf=3, min\_samples\_split=8) \\
RF & MobileMDE & 6 & RandomForestClassifier(max\_depth=80, max\_features=2, min\_samples\_leaf=3, min\_samples\_split=8) \\
RF & MobileMDE & 7 & RandomForestClassifier(max\_depth=80, max\_features=2, min\_samples\_leaf=3, min\_samples\_split=8) \\
RF & MobileMDE & 8 & RandomForestClassifier(max\_depth=80, max\_features=2, min\_samples\_leaf=3, min\_samples\_split=8) \\
RF & MobileMDE & 9 & RandomForestClassifier(max\_depth=80, max\_features=2, min\_samples\_leaf=3, min\_samples\_split=8) \\
RF & MPM4CPS & 0 & RandomForestClassifier(max\_depth=90, max\_features=2, min\_samples\_leaf=3, min\_samples\_split=10) \\
RF & MPM4CPS & 1 & RandomForestClassifier(max\_depth=100, max\_features=3, min\_samples\_leaf=5, min\_samples\_split=10) \\
RF & MPM4CPS & 2 & RandomForestClassifier(max\_depth=80, max\_features=3, min\_samples\_leaf=4, min\_samples\_split=12) \\
RF & MPM4CPS & 3 & RandomForestClassifier(max\_depth=110, max\_features=2, min\_samples\_leaf=3, min\_samples\_split=8) \\
RF & MPM4CPS & 4 & RandomForestClassifier(max\_depth=100, max\_features=3, min\_samples\_leaf=4, min\_samples\_split=8) \\
RF & MPM4CPS & 5 & RandomForestClassifier(max\_depth=80, max\_features=3, min\_samples\_leaf=3, min\_samples\_split=8) \\
RF & MPM4CPS & 6 & RandomForestClassifier(max\_depth=90, max\_features=3, min\_samples\_leaf=3, min\_samples\_split=10, n\_estimators=200) \\
RF & MPM4CPS & 7 & RandomForestClassifier(max\_depth=80, max\_features=3, min\_samples\_leaf=3, min\_samples\_split=8) \\
RF & MPM4CPS & 8 & RandomForestClassifier(max\_depth=110, max\_features=3, min\_samples\_leaf=4, min\_samples\_split=10) \\
RF & MPM4CPS & 9 & RandomForestClassifier(max\_depth=110, max\_features=3, min\_samples\_leaf=3, min\_samples\_split=10, n\_estimators=200) \\
RF & RL4SE &   0 & RandomForestClassifier(max\_depth=80, max\_features=2, min\_samples\_leaf=3, min\_samples\_split=8) \\
RF & RL4SE &   1 & RandomForestClassifier(max\_depth=80, max\_features=2, min\_samples\_leaf=3, min\_samples\_split=8) \\
RF & RL4SE &   2 & RandomForestClassifier(max\_depth=80, max\_features=2, min\_samples\_leaf=3, min\_samples\_split=8) \\
RF & RL4SE &   3 & RandomForestClassifier(max\_depth=80, max\_features=2, min\_samples\_leaf=3, min\_samples\_split=8) \\
RF & RL4SE &   4 & RandomForestClassifier(max\_depth=80, max\_features=2, min\_samples\_leaf=3, min\_samples\_split=8) \\
RF & RL4SE &   5 & RandomForestClassifier(max\_depth=80, max\_features=2, min\_samples\_leaf=3, min\_samples\_split=8) \\
RF & RL4SE &   6 & RandomForestClassifier(max\_depth=80, max\_features=2, min\_samples\_leaf=3, min\_samples\_split=8) \\
RF & RL4SE &   7 & RandomForestClassifier(max\_depth=80, max\_features=2, min\_samples\_leaf=3, min\_samples\_split=8) \\
RF & RL4SE &   8 & RandomForestClassifier(max\_depth=80, max\_features=2, min\_samples\_leaf=3, min\_samples\_split=8) \\
RF & RL4SE &   9 & RandomForestClassifier(max\_depth=80, max\_features=2, min\_samples\_leaf=3, min\_samples\_split=8) \\
RF & UpdateCollabMDE & 0 & RandomForestClassifier(max\_depth=80, max\_features=2, min\_samples\_leaf=3, min\_samples\_split=8) \\
RF & UpdateCollabMDE & 1 & RandomForestClassifier(max\_depth=80, max\_features=2, min\_samples\_leaf=3, min\_samples\_split=8) \\
RF & UpdateCollabMDE & 2 & RandomForestClassifier(max\_depth=80, max\_features=2, min\_samples\_leaf=3, min\_samples\_split=8) \\
RF & UpdateCollabMDE & 3 & RandomForestClassifier(max\_depth=80, max\_features=2, min\_samples\_leaf=3, min\_samples\_split=8) \\
RF & UpdateCollabMDE & 4 & RandomForestClassifier(max\_depth=80, max\_features=2, min\_samples\_leaf=3, min\_samples\_split=8) \\
RF & UpdateCollabMDE & 5 & RandomForestClassifier(max\_depth=80, max\_features=2, min\_samples\_leaf=3, min\_samples\_split=8) \\
RF & UpdateCollabMDE & 6 & RandomForestClassifier(max\_depth=80, max\_features=2, min\_samples\_leaf=3, min\_samples\_split=8) \\
RF & UpdateCollabMDE & 7 & RandomForestClassifier(max\_depth=80, max\_features=2, min\_samples\_leaf=3, min\_samples\_split=8) \\
RF & UpdateCollabMDE & 8 & RandomForestClassifier(max\_depth=80, max\_features=2, min\_samples\_leaf=3, min\_samples\_split=8) \\
RF & UpdateCollabMDE & 9 & RandomForestClassifier(max\_depth=80, max\_features=2, min\_samples\_leaf=3, min\_samples\_split=8) \\
SVC & DSMLCompo & 0 & SVC(C=0.001, gamma=10.0, kernel='poly', probability=True) \\
SVC & DSMLCompo & 1 & SVC(C=0.001, gamma=10.0, kernel='poly', probability=True) \\
SVC & DSMLCompo & 2 & SVC(C=0.001, gamma=10.0, kernel='poly', probability=True) \\
SVC & DSMLCompo & 3 & SVC(C=0.01, gamma=1.0, kernel='poly', probability=True) \\
SVC & DSMLCompo & 4 & SVC(C=0.001, gamma=10.0, kernel='poly', probability=True) \\
SVC & DSMLCompo & 5 & SVC(C=100.0, gamma='auto', kernel='linear', probability=True) \\
SVC & DSMLCompo & 6 & SVC(C=100.0, gamma='auto', kernel='linear', probability=True) \\
SVC & DSMLCompo & 7 & SVC(C=100.0, gamma='auto', kernel='linear', probability=True) \\
SVC & DSMLCompo & 8 & SVC(C=1000.0, gamma='auto', kernel='linear', probability=True) \\
SVC & DSMLCompo & 9 & SVC(C=1000.0, gamma='auto', kernel='linear', probability=True) \\
SVC & MobileMDE & 0 & SVC(C=0.001, gamma=1.0, kernel='poly', probability=True) \\
SVC & MobileMDE & 1 & SVC(C=0.001, gamma=1.0, kernel='poly', probability=True) \\
SVC & MobileMDE & 2 & SVC(C=10.0, gamma='auto', kernel='linear', probability=True) \\
SVC & MobileMDE & 3 & SVC(C=0.001, gamma=1.0, kernel='poly', probability=True) \\
SVC & MobileMDE & 4 & SVC(C=1000.0, gamma='auto', kernel='sigmoid', probability=True) \\
SVC & MobileMDE & 5 & SVC(C=1000.0, gamma=0.1, kernel='sigmoid', probability=True) \\
SVC & MobileMDE & 6 & SVC(C=10.0, gamma='auto', kernel='linear', probability=True) \\
SVC & MobileMDE & 7 & SVC(C=1000.0, gamma=0.1, kernel='sigmoid', probability=True) \\
SVC & MobileMDE & 8 & SVC(C=0.001, gamma=1.0, kernel='poly', probability=True) \\
SVC & MobileMDE & 9 & SVC(C=0.001, gamma='auto', kernel='linear', probability=True) \\
SVC & MPM4CPS & 0 & SVC(C=0.001, gamma=1.0, kernel='poly', probability=True) \\
SVC & MPM4CPS & 1 & SVC(C=0.001, gamma=1.0, kernel='poly', probability=True) \\
SVC & MPM4CPS & 2 & SVC(C=10.0, gamma='auto', kernel='linear', probability=True) \\
SVC & MPM4CPS & 3 & SVC(C=0.01, gamma=1.0, kernel='poly', probability=True) \\
SVC & MPM4CPS & 4 & SVC(C=100.0, gamma='auto', kernel='sigmoid', probability=True) \\
SVC & MPM4CPS & 5 & SVC(C=10.0, gamma='auto', kernel='linear', probability=True) \\
SVC & MPM4CPS & 6 & SVC(C=0.001, gamma=1.0, kernel='poly', probability=True) \\
SVC & MPM4CPS & 7 & SVC(C=10.0, gamma=0.1, probability=True) \\
SVC & MPM4CPS & 8 & SVC(C=1000.0, gamma='auto', kernel='sigmoid', probability=True) \\
SVC & MPM4CPS & 9 & SVC(C=1000.0, gamma=0.1, kernel='sigmoid', probability=True) \\
SVC & RL4SE &   0 & SVC(C=1000.0, gamma='auto', probability=True) \\
SVC & RL4SE &   1 & SVC(C=1000.0, gamma='auto', probability=True) \\
SVC & RL4SE &   2 & SVC(C=1000.0, gamma='auto', probability=True) \\
SVC & RL4SE &   3 & SVC(C=100.0, gamma='auto', kernel='linear', probability=True) \\
SVC & RL4SE &   4 & SVC(C=100.0, gamma='auto', probability=True) \\
SVC & RL4SE &   5 & SVC(C=1000.0, gamma='auto', kernel='sigmoid', probability=True) \\
SVC & RL4SE &   6 & SVC(C=10.0, gamma='auto', kernel='linear', probability=True) \\
SVC & RL4SE &   7 & SVC(C=10.0, gamma=0.1, probability=True) \\
SVC & RL4SE &   8 & SVC(C=10.0, gamma='auto', kernel='linear', probability=True) \\
SVC & RL4SE &   9 & SVC(C=0.001, gamma=10.0, kernel='poly', probability=True) \\
SVC & UpdateCollabMDE & 0 & SVC(C=1000.0, gamma='auto', probability=True) \\
SVC & UpdateCollabMDE & 1 & SVC(C=1000.0, gamma=0.1, kernel='sigmoid', probability=True) \\
SVC & UpdateCollabMDE & 2 & SVC(C=0.001, gamma='auto', kernel='linear', probability=True) \\
SVC & UpdateCollabMDE & 3 & SVC(C=100.0, gamma=0.1, kernel='sigmoid', probability=True) \\
SVC & UpdateCollabMDE & 4 & SVC(C=1000.0, gamma=0.1, kernel='sigmoid', probability=True) \\
SVC & UpdateCollabMDE & 5 & SVC(C=1000.0, gamma=0.1, kernel='sigmoid', probability=True) \\
SVC & UpdateCollabMDE & 6 & SVC(gamma='auto', kernel='linear', probability=True) \\
SVC & UpdateCollabMDE & 7 & SVC(C=10.0, gamma='auto', kernel='linear', probability=True) \\
SVC & UpdateCollabMDE & 8 & SVC(C=10.0, gamma='auto', kernel='linear', probability=True) \\
SVC & UpdateCollabMDE & 9 & SVC(C=10.0, gamma='auto', kernel='linear', probability=True) \\
\end{longtblr}

%% file: sections/appendices/consistency-full.tex
\section{Consistency analysis data}\label{app:consistency}

\begin{table*}[h!]
\centering
\caption{Moment statistics of derived metrics in the RL4SE dataset (N=10). Bold is best.}
\label{tab:consistency-rl4se-full}
  \begin{tabular}{@{}llccccc@{}}
    \toprule
    
    \multirow{8}{*}{bAcc} &
    \multicolumn{1}{c}{\textbf{Model}} &
    \multicolumn{1}{c}{\textbf{Mean}} &
    \multicolumn{1}{c}{\textbf{Median}} &
    \multicolumn{1}{c}{\textbf{Std. dev.}} &
    \multicolumn{1}{c}{\textbf{IQR}} &
    \multicolumn{1}{c}{\textbf{Kurtosis}} \\
    \cmidrule{2-7}
    &Random                        & 0.506 & 0.504 & 0.019 & 0.026 & -0.023 \\
    \cmidrule{2-7}
    &Logistic Regression           & 0.723 & 0.735 & 0.044 & 0.065 &  0.899 \\
    &Complement Naive Bayes        & 0.727 & 0.722 & 0.037 & 0.059 & -0.702 \\
    &Support Vector Classification & 0.726 & 0.722 & 0.038 & 0.044 & 1.143 \\
    &Random Forest                 & 0.719 & 0.725 & 0.025 & 0.033 & 0.669 \\
    \cmidrule{2-7}
    &GPT-3.5                       & 0.755 & 0.753 & \textbf{0.004} & \textbf{0.008} & \textbf{-1.322} \\
    \bottomrule\\

    \multirow{8}{*}{F2} &
    \multicolumn{1}{c}{\textbf{Model}} &
    \multicolumn{1}{c}{\textbf{Mean}} &
    \multicolumn{1}{c}{\textbf{Median}} &
    \multicolumn{1}{c}{\textbf{Std. dev.}} &
    \multicolumn{1}{c}{\textbf{IQR}} &
    \multicolumn{1}{c}{\textbf{Kurtosis}} \\
    \cmidrule{2-7}
    &Random                        & 0.262 & 0.258 & 0.018 & 0.027 & -0.037 \\
    \cmidrule{2-7}
    &Logistic Regression           & 0.486 & 0.499 & 0.060 & 0.089 & -0.036 \\
    &Complement Naive Bayes        & 0.483 & 0.479 & 0.049 & 0.072 & -0.271 \\
    &Support Vector Classification & 0.485 & 0.466 & 0.058 & 0.069 & 1.294 \\
    &Random Forest                 & 0.471 & 0.477 & 0.036 & 0.047 & 1.891 \\
    \cmidrule{2-7}
    &GPT-3.5                       & 0.505 & 0.503 & \textbf{0.004} & \textbf{0.009} & \textbf{-1.231} \\
    \bottomrule \\

    \multirow{8}{*}{MCC} &
    \multicolumn{1}{c}{\textbf{Model}} &
    \multicolumn{1}{c}{\textbf{Mean}} &
    \multicolumn{1}{c}{\textbf{Median}} &
    \multicolumn{1}{c}{\textbf{Std. dev.}} &
    \multicolumn{1}{c}{\textbf{IQR}} &
    \multicolumn{1}{c}{\textbf{Kurtosis}} \\
    \cmidrule{2-7}
    &Random                        & 0.503 & 0.502 & 0.011 & 0.015 & -0.028 \\
    \midrule
    &Logistic Regression           & 0.660 & 0.663 & 0.046 & 0.067 & 0.085 \\
    &Complement Naive Bayes        & 0.648 & 0.649 & 0.031 & 0.053 & -0.375 \\
    &Support Vector Classification & 0.649 & 0.629 & 0.048 & 0.050 & 1.723 \\
    &Random Forest                 & 0.641 & 0.635 & 0.028 & 0.044 & 0.524 \\
    \midrule
    &GPT-3.5                       & 0.649 & 0.648 & \textbf{0.002} & \textbf{0.005} & \textbf{-1.236} \\
    \bottomrule
    
\end{tabular}
\end{table*}

%% file: sections/appendices/significance.tex
\section{Significance analysis data}\label{app:significance}
\begin{table}[!htbp]
  \caption{\textit{p}-values of the statistical tests comparing ChatGPT with the other classifiers in the RL4SE dataset}
  \label{tab:significance-rl4se}
  \begin{tabular}{lccccccc}
    \toprule
     & \textbf{Rec} & \textbf{Prec} & \textbf{Spec} & \textbf{NPV} & \textbf{bAcc}  & \textbf{F2}  & \textbf{MCC} \\
    \toprule
    Random                          & $\mathbf{0.000^{*}}$ & $\mathbf{0.002^{*}}$ & $\mathbf{0.001^{*}}$ & $\mathbf{0.000^{*}}$ & $\mathbf{0.000^{*}}$ & $\mathbf{0.000^{*}}$ & $\mathbf{0.000^{*}}$ \\
    \midrule
    Logistic Regression             & $\mathbf{0.000^{*}}$ & $0.193$ & $0.315$ & $\mathbf{0.000^{*}}$ & $0.542$ & $1.000$ & $1.000$ \\
    Complement Naive Bayes          & $0.098$ & $1.000$ & $1.000$ & $0.055$ & $1.000$ & $1.000$ & $1.000$ \\
    Support Vector Classification   & $0.555$ & $1.000$ & $1.000$ & $0.121$ & $0.989$ & $1.000$ & $1.000$ \\
    Random Forest                   & $\mathbf{0.009^{*}}$ & $1.000$ & $1.000$ & $\mathbf{0.001^{*}}$ & $0.195$ & $1.000$ & $1.000$ \\
    \bottomrule
  \end{tabular}
  \vspace{1pt}
  \caption*{$^{*}$: significant at $\alpha=0.05$. Bold indicates ChatGPT performs significantly better.}
\end{table}
\begin{table}[!htbp]
  \caption{\textit{p}-values of the statistical tests comparing ChatGPT with the other classifiers in the DSMLCompo dataset}
  \label{tab:significance-dsmlcompo}
  \begin{tabular}{lccccccc}
    \toprule
     & \textbf{Rec} & \textbf{Prec} & \textbf{Spec} & \textbf{NPV} & \textbf{bAcc}  & \textbf{F2}  & \textbf{MCC} \\
    \toprule
    Random                          & $\mathbf{0.000^{*}}$ & $\mathbf{0.000^{*}}$ & $\mathbf{0.000^{*}}$ & $\mathbf{0.000^{*}}$ & $\mathbf{0.000^{*}}$ & $\mathbf{0.000^{*}}$ & $\mathbf{0.000^{*}}$ \\
    \midrule
    Logistic Regression             & $0.475$       & $\mathbf{0.022^{*}}$ & $1.000$ & $\mathbf{0.003^{*}}$  & $\mathbf{0.000^{*}}$ & $\mathbf{0.000^{*}}$ & $\mathbf{0.000^{*}}$ \\
    Complement Naive Bayes          & $0.679$       & $0.169$     & $1.000$ & $\mathbf{0.016^{*}}$   & $\mathbf{0.000^{*}}$ & $\mathbf{0.000^{*}}$ & $\mathbf{0.000^{*}}$ \\
    Support Vector Classification   & $\mathbf{0.004^{*}}$   & $0.053$     & $1.000$ & $\mathbf{0.000^{*}}$ & $\mathbf{0.000^{*}}$ & $\mathbf{0.000^{*}}$ & $\mathbf{0.000^{*}}$ \\
    Random Forest                   & $\mathbf{0.000^{*}}$ & $1.000$     & $1.000$ & $\mathbf{0.000^{*}}$ & $\mathbf{0.000^{*}}$ & $\mathbf{0.000^{*}}$ & $\mathbf{0.000^{*}}$ \\
    \bottomrule
  \end{tabular}
  \vspace{1pt}
  \caption*{$^{*}$: significant at $\alpha=0.05$. Bold indicates ChatGPT performs significantly better.}
\end{table}